\documentclass[manuscript]{acmart}
\AtBeginDocument{%
  \providecommand\BibTeX{{%
    \normalfont B\kern-0.5em{\scshape i\kern-0.25em b}\kern-0.8em\TeX}}}

\setcopyright{acmcopyright}
\copyrightyear{2024} 
\acmYear{2024} 
\setcopyright{rightsretained} 
\acmConference[CHI '24]{Proceedings of the CHI Conference on Human Factors in Computing Systems}{May 11--16, 2024}{Honolulu, HI, USA}
\acmBooktitle{Proceedings of the CHI Conference on Human Factors in Computing Systems (CHI '24), May 11--16, 2024, Honolulu, HI, USA}
\acmDOI{10.1145/3613904.3641902}
\acmISBN{979-8-4007-0330-0/24/05}

\usepackage{csquotes}
\usepackage{multirow}
\usepackage{acmart-taps}

\definecolor{darkgreen}{RGB}{26, 102, 46}

\begin{document}

\title{(Beyond) Reasonable Doubt: Challenges that Public Defenders Face in Scrutinizing AI in Court}

\author{Angela Jin}
\email{angelacjin@berkeley.edu}
\affiliation{%
  \institution{University of California, Berkeley}
  \city{Berkeley}
  \country{U.S.A.}}

\author{Niloufar Salehi}
\email{salehi@ischool.berkeley.edu}
\affiliation{%
  \institution{University of California, Berkeley}
  \city{Berkeley}
  \country{U.S.A.}}

\renewcommand{\shortauthors}{Jin and Salehi}


\begin{abstract}
Accountable use of AI systems in high-stakes settings relies on making systems contestable. In this paper we study efforts to contest AI systems in practice by studying how public defenders scrutinize AI in court. We present findings from interviews with 17 people in the U.S. public defense community to understand their perceptions of and experiences scrutinizing computational forensic software (CFS) — automated decision systems that the government uses to convict and incarcerate, such as facial recognition, gunshot detection, and probabilistic genotyping tools. We find that our participants faced challenges assessing and contesting CFS reliability due to difficulties (a) navigating how CFS is developed and used, (b) overcoming judges and jurors’ non-critical perceptions of CFS, and (c) gathering CFS expertise. To conclude, we provide recommendations that center the technical, social, and institutional context to better position interventions such as performance evaluations to support contestability in practice. 
\end{abstract}





\maketitle

\section{Introduction}

Making systems contestable --- i.e., open to scrutiny and disagreement --- presents a promising approach to ensuring responsible and accountable use of automated decision systems in domains such as criminal law, healthcare, and social work~\cite{kluttz2022shaping, vaccaro2019contestability, alfrink2022contestable, kuo2023understanding, wexler2017computer}. This is particularly important for stochastic systems that have variable accuracy, such as those using machine learning~\cite{bansal2019beyond}. Performance evaluations of these systems play an important role in supporting contestability, as evaluation study procedures and results can provide crucial insights into system reliability for those seeking to scrutinize and challenge algorithmically-driven decisions. Yet, recent work has revealed numerous ways in which real-world practices in evaluation design, documentation, and reporting fall short of these ideals. For example, test datasets may fail to accurately represent real-world uses~\cite{raji2021ai, kuo2023understanding, taori2020measuring}, performance metrics may misalign with users' or other downstream stakeholders’ perceptions of success~\cite{patton2020contextual, kawakami2022care, kuo2023understanding}, and published documents describing evaluations may omit important details about procedures and results~\cite{metcalf2023taking, wu2021medical, butler2021dna}. These practices present challenges for multiple downstream stakeholders (e.g., users, decision subjects, regulators) and can be especially restricting for decision subjects who may be harmed by incorrect, uncontested system outputs. In this paper, we focus on the experiences of the advocates representing decision subjects. We ask: \textit{What challenges do public defenders face when scrutinizing the government's\footnote{We use ``government'' to refer to the prosecution, who represents the government in criminal cases.} use of automated decision systems in the U.S. criminal legal system?} 

We specifically focus on the U.S. criminal legal system's increasing reliance on \textit{computational forensic software (CFS)} --- one common type of automated decision system that the government uses to convict and incarcerate, such as facial recognition, gunshot detection, and probabilistic genotyping tools. We investigate how public defenders assess and contest the reliability of CFS and the government's use of CFS. Amidst growing calls for laws and policies that shape performance evaluations in the U.S. criminal legal system~\cite{takano2021, garrett2020error, garvie2022forensic, butler2021dna, lander2016forensic}, our work seeks to contribute insights that center the needs and experiences of public defenders~\cite{warren2022trial}\footnote{As emphasized in recent work surveying existing approaches to participatory design of AI systems, while public defenders represent impacted communities subject to decisions made using CFS, their perspectives should not be interpreted as proxies for those of indigent defendants~\cite{delgado2023participatory}. We highlight that our exploration of public defenders' needs can additionally serve as an initial step towards this latter goal of understanding decision subjects' needs, which we highlight as an area for future work.}.

We conducted 17 semi-structured interviews with individuals from the U.S. public defense community, focusing on public defenders and those who work with public defenders on the technology-related aspects of cases (e.g., other lawyers with expertise in CFS, individuals with science or technology expertise working in public defense offices). In our interviews, we sought to learn about participants' past encounters with CFS in casework, how they attempted to assess and contest reliability, and challenges they faced in doing so.

Our findings reveal a wide range of technical, social, and institutional challenges that public defenders face in assessing and contesting CFS reliability. We organize our presentation of these findings in three categories representing challenges that public defenders face in (1) navigating CFS developers and users' practices and policies, (2) overcoming judges and jurors' non-critical perceptions of CFS, and (3) gathering CFS expertise. First, our participants often felt that software developers and users' practices and policies for developing, testing, using, and sharing information about CFS severely constrained their efforts to assess and contest CFS reliability. For example, several participants expressed frustration at prosecutors and software companies withdrawing CFS evidence when judges granted defenders opportunities to rigorously scrutinize the CFS tool in the case at hand. Second, our participants highlighted how judges and jurors’ non-critical perceptions of CFS could additionally constrain public defenders’ efforts to assess and contest reliability. To grapple with these challenges, the defenders we spoke to relied heavily on expert witnesses and colleagues with relevant CFS expertise. These collaborations helped public defenders identify important information to request during discovery, identify potential flaws in CFS outputs, craft their arguments, and explain the implications of their findings to judges and jurors. However, when seeking outside experts, public defenders faced difficulties finding people with relevant expertise who were available and willing to work with them, and insufficient funding continues to limit defense offices' ability to hire and build in-office expertise.

Based on these findings, we identify opportunities for future work in human-AI interaction to engage with work in public policy and responsible AI, towards ensuring that performance evaluations are effective for advocates seeking to assess and contest automated decision systems when used to make individual decisions. First, we highlight the importance of factors outside the design and communication of performance evaluations, and we discuss the role that HCI can play in overcoming barriers to leveraging performance evaluations as a tool for contesting automated decision systems. Second, engaging public defenders in the design of CFS performance evaluations can provide valuable insights into the reliability of CFS, but addressing barriers to engaging them in evaluation design, and building tools to support their existing needs for collaboration and skill-sharing are crucial. Lastly, we argue that work exploring how decision-makers perceive performance information should explore opportunities to incorporate processes of deliberation, presentation of performance information, and prior beliefs and knowledge of technology and technology users into their study designs.

Taken together, these findings provide insight into advocates' experiences assessing and contesting the reliability of algorithmic systems used to make decisions about the individuals they represent, and contribute implications towards ensuring that performance evaluations effectively support contestability in practice.

\section{Background}
\label{sec:background}

\subsection{Public defense in the U.S.}
The U.S. Constitution guarantees criminal defendants the right to have an attorney assist in their defense\footnote{The Sixth Amendment states that all defendants accused of crimes in the U.S. federal court system have a right to effective assistance by defense counsel. This right was not guaranteed in state courts until Gideon v. Wainwright, 372 U.S. 335 (1963), the landmark U.S. Supreme Court case that extended the Sixth Amendment right to counsel to criminal defendants in state courts.}. For indigent \nobreak defendants who cannot afford to pay for their own lawyer, this right is protected through several different indigent defense services: public defender offices, individually assigned private attorneys, and contract-attorney organizations in state courts; and federal public defender organizations, community defense organizations, and appointed private attorneys in federal courts~\cite{spangenberg1995indigent}. In this work, we focus on full-time public defenders who represent a vast number\footnote{One study found that most defendants charged with felonies --- i.e., violent or serious crimes --- rely on publicly funded attorneys~\cite{harlow2001defense}. Other work has estimated that roughly 5.6 million Americans rely on public defenders~\cite{lynch}.} of individuals who come into contact with the U.S. criminal legal system through both federal and state courts. 

Despite the critical role that public defenders play in protecting the lives and liberties of indigent defendants, public defense offices in the U.S. are often underfunded, and, as a result, lawyers working in these offices are overburdened. For example, one nation-wide study surveying public defender services in the U.S. found that almost 75\% of county-based public defender offices exceeded the maximum recommended limit of cases (400 misdemeanors or 150 felonies per attorney per year) received per attorney~\cite{farole2010county}. A 2019 report detailed that until recently in New Orleans, individual defenders had been forced to handle upward of 19,000 misdemeanor cases in a year, translating into seven minutes per client~\cite{furst2019fair}.

\subsection{Computational forensic software in the U.S. criminal legal system}
In our work, we focus on \textit{computational forensic software (CFS)} ---  automated decision systems that the government increasingly relies on to convict and incarcerate\footnote{The Justice in Forensic Algorithms Act of 2021 defines computational forensic software as ``software that relies on an automated or semiautomated computational process, including one derived from machine learning, statistics, or other data processing or artificial intelligence techniques, to process, analyze, or interpret evidence''~\cite{takano2021}.}. Here, we interpret CFS to include \textit{facial recognition systems} that police use to compare stills from video footage to databases of faces~\cite{garvie2022forensic, facialrecogUse}, \textit{gunshot detection systems} that police use to detect and locate gunshot sounds~\cite{shotspotterUse}, \textit{probabilistic genotyping software} that forensic laboratories use to interpret DNA mixtures~\cite{fstUse}, \textit{automated license plate readers}~\cite{eff_alpr}, \textit{automated fingerprint identification systems}~\cite{moses2011automated}, and \textit{toolmark analysis systems} that aim to recognize and compare patterns in marks made by tools and firearms~\cite{petraco2012application}. In the remainder of this section, we illustrate government use of CFS through facial recognition and probabilistic genotyping software (PGS), which we use as case studies to ground our discussions with public defenders and in this paper. We focus on these two types of CFS for three primary reasons. First, facial recognition and PGS systems have been used in the U.S. for over a decade, and both their adoption by government agencies and use in criminal cases continues to grow. Second, a growing body of work in academia, civil society, and government have raised concerns over the reliability of these systems. Lastly, simultaneously considering facial recognition and PGS allows us to explore how their similarities (e.g., stochastic systems with varying performance across settings) and differences (e.g., primary use to produce investigative leads vs. trial evidence) impact public defenders' experiences and practices.

\subsubsection{Facial recognition}
Police departments rely on facial recognition systems to help identify individuals captured in an image by comparing the image against a database of images of faces with known identities. At a high-level, a police officer chooses a probe photo --- a photograph of the unknown subject of interest (e.g., a video still from a surveillance camera). The officer then chooses a database to run the probe photo against, and inputs the photo into the facial recognition software, sometimes editing the probe photo beforehand. After running the search, officer views the list of possible matches output by the software and manually compares the original probe photo and output candidate photos to determine whether or not the system has produced a possible match. Many police departments in the U.S. consider (i.e., document in formal guidance) possible matches produced by facial recognition systems to be investigative leads and not evidence of probable cause to arrest~\cite{garvie2022forensic}. 

\citet{garvie2016perpetual} reported that one Florida sheriff's office began implementing facial recognition in 2001, and by 2016, at least a quarter of the 15,000 state and local law enforcement agencies in the country had access to facial recognition technology. Facial recognition systems have been --- and continue to be --- the subject of immense scrutiny by civil society and academic research in HCI and responsible AI~(e.g., \cite{buolamwini2018gender, aclu_fr_scrutiny, alkhatib2021live}). 

\subsubsection{Probabilistic genotyping software (PGS)}
Forensic laboratories use PGS in forensic DNA analysis to compare the genetic profile of a person of interest against a DNA sample obtained from the crime scene, typically relying on PGS when the DNA evidence is deemed too complex for manual inspection (e.g., crime scene samples that contain small amounts of degraded DNA from many individuals). PGS systems perform this comparison and compute a likelihood ratio that compares the likelihood of observing the DNA evidence under two competing hypotheses (e.g., the defendant and two unknown others contributed to the DNA mixture vs. three unknown individuals who are not the defendant contributed to the DNA mixture)~\cite{coble2019probabilistic}. At its core, PGS rely on stochastic methods and implement a likelihood ratio test --- fundamental components of many statistical decision-making systems.

In practice, a lab analyst begins their analysis of a forensic DNA sample by preparing the crime scene sample and a comparison profile (typically from the defendant), and input both profiles into the PGS system. The analyst will then specify additional input parameters (e.g., the analyst's inference for the number of contributors to the mixture, the two hypotheses to use for calculating the likelihood ratio) and run the software to produce a likelihood ratio for the defendant. These steps will then result in a report that the prosecution may seek to introduce at trial as incriminatory evidence.

In a recently published comprehensive review of PGS,~\citet{butler2021dna} reported uses of PGS in forensic casework in the U.S. starting as early as 2011. According to the developers of one PGS system, 80\% of certified U.S. labs are currently using the system in casework, are in the process of validating the software for use in casework, or have purchased the software but have not yet started the validation process~\cite{strmixUse}. Yet, the reliability of PGS systems on the complex DNA samples they are often used to analyze is the subject of much debate~\cite{butler2021dna, lander2016forensic, matthews2019right, matthews2020trusted}.

\section{Related Work}
\label{sec:related-work}
We draw on prior work studying the technology-related needs and challenges that public defenders face in criminal cases~\cite{gaub2019understanding, warren2022trial, wilson2023digital}. This work has examined how government use of body-worn cameras~\cite{gaub2019understanding}, surveillance data~\cite{warren2022trial}, and digital evidence more broadly~\cite{wilson2023digital} impact public defense practice. Building on this work, we focus on computational decision-making systems and U.S. public defenders' experiences assessing and contesting the reliability of their outputs. Our investigation of defenders' experiences assessing and contesting CFS reliability also builds on prior work on (1) contesting automated decision systems and (2) designing and communicating performance evaluations.

\subsection{Contesting automated decision systems}
Prior work at the intersection of law, technology, and HCI highlights the importance of ensuring that automated decision systems are contestable (e.g.,~\cite{hirsch2017designing, yurrita2023disentangling, metcalf2023taking}). Theoretical arguments have illustrated how contestability can help promote fairness and accountability in decision-making~\cite{kaminski2021right, hildebrandt2019privacy, hirsch2017designing}, protect individual rights~\cite{kaminski2021right, hildebrandt2019privacy, almada2019human}, and preserve human dignity and autonomy~\cite{kluttz2022shaping}. Recent work has also empirically demonstrated that contestability shapes peoples' perceptions of fairness~\cite{yurrita2023disentangling}, and has shown how the \textit{design} of appeals processes can impact peoples' perceptions of fairness~\cite{lyons2022whats, vaccaro2020end}.

Building on these justifications, a growing body of work in HCI seeks to articulate what contestability requires~\cite{sarra2020put, almada2019human, vaccaro2019contestability, hirsch2017designing}. This body of literature proposes information (e.g., performance evaluation results, documents describing system development and use) and practices (e.g., human intervention, deep engagement, agonistic debate, dialectical exchange) that support contestation. In doing so, this literature collectively characterizes different types of contestation that vary along multiple dimensions such as \textit{who contests} (e.g., developers, users, decision subjects, third parties representing decision subjects), \textit{what's being contested} (e.g., individual algorithm outputs, individual decisions made using algorithm output, the system as a whole), and \textit{when contestation occurs} (e.g., throughout development, during decision-making, after a decision has been put into force)~\cite{alfrink2022contestable}. In turn, each type of contestation implicates specific information needs and processes depending on the stakeholders, system components, and stages involved~\cite{suresh2021beyond}. For example, the confidence level associated with an algorithm's output may be useful for system users \cite{hirsch2017designing}, and decision subjects and third parties representing decision subjects  \cite{alfrink2022contestable}. Meanwhile, technical, organizational, and social accounts of how a decision was made may primarily be useful for decision subjects and their advocates \cite{alfrink2022contestable}. 

However, implementing these practices and ensuring their effectiveness in practice can be challenging due to real-world needs and constraints. In a study using speculative design and semi-structured interviews with civil servants who work with a publicly-deployed AI system in Amsterdam, \citet{alfrink2023contestable} find that implementing contestability in public AI in practice can be challenging for numerous reasons such as citizen capacity (e.g., it can be hard for citizens to understand metrics used for evaluating model performance). Studying treatment of liability and redress cases in U.S. civil courts, \citet{metcalf2023taking} reveal how plaintiffs may face challenges contesting algorithmic harms in practice, due to lack of available documentation, difficulties in convincing courts to hear their case, and lack of recognized expertise. 
We seek to expand these efforts to empirically ground contestation in real-world settings~\cite{metcalf2023taking, alfrink2023contestable} by studying contestation of individual decisions by advocates representing decision subjects (specifically public defenders) within the U.S. criminal legal system. Within this context, we specifically focus on performance evaluations, as several groups have pointed to measures of uncertainty about the system's output (e.g., confidence in output, performance, probability of alternative outcomes) and documentation of system performance as concrete artifacts that can support contestation~\cite{kluttz2022shaping, hirsch2017designing, ploug2020four, alfrink2022contestable, alfrink2023contestable}. We discuss the details of designing and communicating performance evaluations for real-world contexts in the next subsection.

\subsection{Designing and communicating performance evaluations}
A growing body of work at the intersection of algorithmic accountability, human-AI collaboration, and explainable AI explores approaches to designing and communicating performance evaluations such that they are are useful for downstream users and other stakeholders.

\subsubsection{Designing performance evaluations} Prior work has identified how algorithmic systems' performance on static benchmark datasets may fall short of end-users' needs. For instance, test inputs may not be sufficiently representative of real-world settings~\cite{raji2021ai, kuo2023understanding}, and performance metrics may not align with users' preferences and perceptions of ideal model performance~\cite{patton2020contextual, kuo2023understanding, kawakami2022care}. To address this gap, a growing body of work in HCI aims to design performance evaluations grounded in downstream deployment contexts and the needs and goals of downstream stakeholders (e.g.,~\cite{suresh2023kaleidoscope, liao2023rethinking, suresh2022towards, cabrera2023zeno}). This typically involves exploring users' domain-specific information needs~\cite{cai2019hello, kawakami2022improving}, directly working with downstream stakeholders to collaboratively design evaluation datasets and metrics~\cite{suresh2022towards}, and designing tools that allow users to specify their own test datasets and performance metrics~\cite{suresh2023kaleidoscope, devos2022toward, lam2022end, cabrera2023zeno, deng2023understanding}. This ``participatory turn''~\cite{delgado2023participatory} in the design of performance evaluations highlights the importance and strength of centering the experiential and domain expertise of stakeholders in downstream deployment contexts.

In the healthcare context,~\citet{cai2019hello} reveal that pathologists desire not only the AI assistant's overall performance, but also its performance under specific conditions such as well-known edge cases and its ``medical point-of-view'' (e.g., the extent to which the AI system tends to be more liberal or conservative in its diagnoses). Similarly, in their study of child welfare workers' use of algorithmic risk prediction models, \citet{kawakami2022improving} find that workers engaged in ``everyday algorithm auditing''~\cite{devos2022toward} to understand model behaviors and limitations, sometimes running counterfactual scenarios through the model to understand how changing a factor impacted the algorithm's output. We build on this work by focusing on the evaluation needs and goals of public defenders who represent criminal defendants subject to CFS outputs.

\subsubsection{Communicating performance evaluations}
Beyond contextualizing the design of performance evaluations, prior work has also leveraged HCI techniques to explore approaches to communicating evaluation results to various stakeholders to help them determine whether and when to trust system outputs. This body of work typically seeks to understand how users perceive, understand, and utilize performance information such as accuracy and uncertainty~\cite{yin2019understanding, prabhudesai2023understanding, lai2019human, zhang2020effect, yang2023subjective, bansal2019beyond}. 

Through a large-scale laboratory experiment studying the impact of stated and observed accuracy on laypeople's trust in the model,~\citet{yin2019understanding} find that both stated and observed accuracy affected participants' trust, and that observed accuracy mediated people's understanding of stated accuracy. \citet{prabhudesai2023understanding} complement this work through a user study presenting uncertainty information to lay decision-makers and find that communicating uncertainty about ML predictions forced users to more critically engage with the system's output. Directly, relevant to our focus on CFS in the U.S. criminal legal system, \citet{garrett2020error} investigate how jurors perceive error rates associated with forensic evidence. Collectively, this body of work suggests a need to understand what specific performance information downstream stakeholders find useful, along with how they understand and make use of it.
\section{Methodology}
\label{sec:methods}

\begin{table*}
\begin{tabular}{ | c | l | }
\hline
 \textbf{Participant ID} & \textbf{Role} \\ 
 \hline\hline
 P1 & Technologist who built software for public defenders \\  
 \hline
 P2 & Scientist helping lawyers with forensics issues in a public defense office \\
 \hline
 P3 & Public defender for over 10 years \\
 \hline
 P4 & Public defender for over 10 years\\
 \hline
 P5 & Public defender for over 20 years \\
 \hline
 P6 & Public defender for over 10 years \\
 \hline
 P7 & Public defender for 2 years \\
 \hline
 P8 & Public defender for over 10 years\\
 \hline
 P9 & Public defender for 9 years \\
 \hline
 P10 & Public defender for a year \\
 \hline
 P11 & Public defender for over 10 years \\
 \hline
 P12 & Public defender for over 20 years \\
 \hline
 P13 & Public defender for over 10 years \\
 \hline
 P14 & Public defender for over 10 years \\
 \hline
 P15 & Technology law expert helping public defenders on CFS-related issues \\
 \hline
 P16* & Public defender for 5 years \\
 \hline
 P17* & Technologist helping lawyers with digital evidence in a public defense office \\
 \hline
\end{tabular}
\caption{Description of each participant. \textit{*P16 and P17 were interviewed jointly.}}
\label{table:participant-info}
\end{table*}

\subsection{Data collection}

\subsubsection{Participant recruitment}
We adopted a mix of convenience and purposive sampling to recruit individuals from the U.S. public defense community. We initially recruited participants through the National Association of Criminal Defense Lawyers (NACDL) mailing list for public defenders and leveraged contacts from past and ongoing collaborations with researchers and practitioners working in criminal defense. Based on these initial contacts, we purposefully sampled for range and diversity asking for introductions to people who ranged in terms of region. In total, we recruited 17 participants. According to the four regions defined by the US Census Bureau\footnote{\url{https://www2.census.gov/geo/pdfs/maps-data/maps/reference/us_regdiv.pdf}}, 9 of our participants work or have worked on cases tried in the Northeast, 5 in the South, 1 in the Midwest, and 5 in the West\footnote{These counts sum up to greater than 17 because some of our participants have worked on cases tried by courts in multiple U.S. regions.}. Table~\ref{table:participant-info} summarizes participants' roles. 

\subsubsection{Interview protocol}
We conducted interviews from December 2022 to August 2023. 15 interviews were conducted over video call (primarily over Zoom, but sometimes over Google Meet or Microsoft Teams, depending on the participant's preferences), and 1 interview was conducted over phone. Most interviews lasted approximately 1 hour, with just one lasting 20 minutes given the participant's time constraints. We offered every participant a \$25 e-gift card of their choice. We recorded the interview as per the participants’ approval, transcribed the recording, and anonymized the transcription. 

Our primary goal in this research was to understand the challenges that public defenders face assessing and contesting reliability of CFS and the government’s use of CFS in criminal trials. To this end, we were also interested in understanding participants’ general perceptions of CFS and perceptions of potential approaches to evaluating the reliability of CFS. We explored these research questions through semi-structured interviews in which we guide the conversation in three parts.

In part 1, to understand participants’ background, general familiarity with CFS, and general perceptions of CFS, we asked questions about their current role, the types of CFS they know of or have encountered in past cases, and their feelings about the use of CFS in the U.S. criminal legal system. We then focused our discussions in parts 2 and 3 on 1-3 specific CFS systems that the participant was most familiar with (i.e.,  had experience litigating against the CFS, or had learned about the CFS through other means), prioritizing facial recognition and probabilistic genotyping when participants were familiar with either or both. In part 2, to understand the challenges participants faced in assessing and contesting CFS reliability, we asked participants to talk about their experiences when the government used CFS in specific cases, with a particular emphasis on the challenges they experienced when gathering information to assess CFS reliability and when contesting reliability. However, due to confidentiality and case sensitivity, public defenders sometimes preferred to talk about their past case experiences more generally. Lastly, in part 3, to understand how participants felt about potential approaches to assessing CFS reliability, we introduced storyboards depicting evaluation approaches inspired by those proposed in recent literature and policies (e.g.,~\cite{abebe2022adversarial, takano2021}) (see Appendix~\ref{appendix}). One storyboard depicted steps that a public defender might take in designing a performance evaluation of a PGS system: adversarially choosing input data, measuring CFS performance on the chosen inputs, and communicating results to courts and jurors (e.g., \cite{abebe2022adversarial}). Another depicted steps inspired by Model Cards~\cite{mitchell2019model}. We then asked follow-up questions such as ``How would you present the results of this test to a judge?'', and ``What challenges, if any, would you anticipate facing if you presented this information to jurors?'' In interviews where participants had past experience litigating against CFS but did not feel comfortable sharing specific case details in the interview, we used these storyboards to discuss hypothetical scenarios in a way that was more grounded and would not have been possible without the use of the storyboards. Sometimes participants were comfortable talking about specific cases without mentioning client details, but had not yet litigated against CFS or had not used the specific approaches presented in the storyboards. In these situations, the storyboards helped connect public defenders' existing knowledge of CFS, legal processes, and legal actors with potential futures proposed in recent literature and policies. Not all interviews involved storyboards, as we sometimes chose to prioritize other points discussed in the interview.

\subsubsection{Limitations and opportunities}
Our semi-structured interview approach helps us elicit illustrative experiences and perspectives, towards our goal of developing an in-depth view of the challenges that public defenders face in scrutinizing CFS in the U.S. criminal legal system. While we strive to gather as diverse of range of a perspective as possible through our recruitment approaches, we caution against assuming that these experiences and perspectives speak for public defenders throughout the country, as technology uses, office norms, and defender perspectives may vary across jurisdictions, public defense offices, and individual public defenders.

We intentionally center public defenders given existing power imbalances in the U.S. criminal legal system and how the system disproportionately impacts low-income and marginalized communities. However, providing a complete picture of the experiences of impacted individuals requires engagement with criminal defendants, along with the broader low-income and marginalized communities that these public defenders serve. In speaking with people who are part of the public defense community, we also omit perspectives of other stakeholders, such as judges, jurors, prosecutors, law enforcement, forensic labs, and CFS developers.

\subsection{Data analysis}
Our interviews yielded 15.3 hours of audio recording, which we analyzed using inductive qualitative analysis, drawing on elements of grounded theory methodology~\cite{charmaz2006constructing}. After anonymizing and fixing transcription errors in each transcript, we conducted line-by-line open coding. We then identified relationships between codes and grouped codes into increasingly abstract themes through an iterative, bottom-up affinity diagramming process~\cite{beyer1999contextual}. In total, this process yielded three levels of themes: 20 first-level themes, 11 second-level themes, and 3 third-level themes. Examples of first-level themes include ``CSI effect of DNA'', ``judge siding with prosecution'', and ``lab communicating with prosecution''. 

\subsection{Positionality}
Both authors are researchers trained in the United States in fields of Human-Computer Interaction and Artificial Intelligence. The questions we ask and our research approach are shaped by our prior and ongoing collaborations with those who work within the U.S. criminal legal system, but neither of us has professional or personal experience in the U.S. criminal legal system. 

We acknowledge that HCI research framed as participatory can take up participants' time and energy while providing no tangible benefits for participants~\cite{pierre2021getting}. This is especially of concern for us since public defenders must work within a system that disproportionately targets and systemically disadvantages their clients. As a result, public defenders are often overburdened with extreme caseloads while having limited resources to tackle them. With all of this in mind, we plan to share summarized insights with our participants and translate our research findings to policy insights and proposals. We intend to continue our collaboration with public defenders and those who work with them.
\section{Results}
\label{sec:results}
We found that public defenders face three primary challenges when scrutinizing the government's use of CFS. First, public defenders often felt that software users and developers' actions constrained their ability to assess and contest the reliability of the CFS outputs. For instance, decisions that forensic labs and CFS developers made when testing these tools could constrain the extent to which defenders were able to assess reliability. Similarly, prosecutors could withdraw evidence or offer plea deals to prevent defenders from scrutinizing and contesting CFS outputs. 

Second, public defenders additionally needed to convince judges and jurors that the software output was unreliable to ensure that these decision-makers took appropriate actions (e.g., that judges declare the CFS output inadmissible, or that jurors determine that the CFS output does not contribute to their `beyond reasonable doubt' determination). Yet our participants often grappled with judges and jurors' non-critical perceptions of technology and CFS users. 

Third, to overcome these challenges, our participants relied heavily on expert witnesses to help assess CFS reliability, craft arguments, and testify in court on their behalf. However, our participants experienced difficulties in locating experts available and willing to work with them. Building up in-house expertise helped alleviate these challenges, but doing so remains a costly endeavor that most public defense offices are unable to achieve due to insufficient funding. In this section, we explore each in turn.

\subsection{Navigating policies and practices of CFS users and developers} 
Many of our participants described how the policies and practices of CFS users (e.g., prosecutors, forensic labs, police departments) and developers (e.g., private companies) could limit the extent to which public defenders were able to gather information to assess software reliability. Even if defenders gathered sufficient information to raise concerns with CFS reliability, prosecutors and companies could withdraw evidence or settle the case via a plea deal, consequently precluding defense from contesting software reliability in the case at hand. Our participants reported feeling frustrated by the difficulties obtaining information and opportunities to assess and contest reliability, as it could severely hinder the representation they were able to provide to their current and future clients.

\subsubsection{Not knowing about CFS use} \label{sec:challenge:identifying-sw-use} In our study and in past work, public defenders commonly felt the need to scrutinize every piece of information used to incriminate their clients~\cite{warren2022trial}. For instance, if police officers relied on facial recognition systems to arrest their client\footnote{\citet{garvie2022forensic} finds that, while police department policies and guidance dictates that a facial recognition identification is insufficient to demonstrate probable cause to make an arrest, there have been several instances where police relied heavily, if not exclusively, on facial recognition results to make arrests.}, defenders would want to assess the reliability of the facial recognition system used and the process through which the officer used the CFS. However, we find that practices and policies of police departments and forensic labs sometimes prevented public defenders from knowing that a CFS tool was used to arrest or incriminate their client. Without knowing this information, public defenders could not scrutinize or contest the software output that might play a crucial role in their client's case. Our participants illustrated two situations in which this challenge could arise in individual cases: when the government proactively chose to not disclose CFS use, and when the government did not explicitly mention CFS use and public defenders lacked specific CFS knowledge to identify important keywords in the reports they received. We additionally find that these challenges in individual cases were exacerbated by broader difficulties public defenders faced in keeping up to date with government adoption and use of new CFS.

Several participants felt that police departments were proactively hiding their use of facial recognition [P5, P13, P15, P16, P17]. As P13 illustrated, \textit{``[the police] just hid the fact that they used [facial recognition] completely and created these sort of parallel constructions where you wouldn't understand how they got from no suspect to suspect.''} In the words of P15, the police was only \textit{``gonna tell you that there's an eyewitness identification.''} This proactive nondisclosure discussed aligns with findings from an empirical investigation by~\citet{garvie2022forensic}, which found evidence that at least one police department intentionally hid the use of facial recognition from judicial proceedings. Related work has also found that software companies may create these challenges by forbidding police from disclosing new software~\cite{wexler2018life, joh2017undue}, or by introducing new, undisclosed technologies within existing known technologies (e.g., body camera providers introducing facial recognition systems within the tools they already provide)~\cite{warren2022trial}.

Another reason why CFS use might go undetected was when the public defender lacked the knowledge to identify language in the reports that might have suggested CFS use. Our participants illustrated how these situations may arise in government uses of PGS. Reflecting on their earliest experiences helping out with cases involving PGS, P15 explained: \textit{``[Public defenders] would know DNA was available in their case because they would get a lab report that would say `DNA samples tested', but the public defenders didn't know enough about [PGS] to identify, even when they had it --- like it was in front of them, it wasn't really being hidden. [T]hey didn't know how to translate the words on that report to realize this is a different type of DNA testing than run of the mill gold standard DNA. [\ldots][T]here was some disclosure, but [\ldots] the prosecutor didn't say, `Dear public defender, I'm just letting you know there was software in your case.' And public defenders didn't know what keywords to be looking out for. So they didn't know, oh, this is a case where I have [PGS] and maybe I want to challenge that as a software problem.''} P5's own experiences complemented this finding: \textit{``our [jurisdiction's] laboratory has just gotten and validated [PGS]. That doesn't seem to be showing up in the reports. It's not saying, `On the [PGS] machine we got this kind of a result.'''}

We further illustrate the difficulty of differentiating between statistics output by PGS and traditional DNA analysis with the following example. Traditional DNA analysis may produce results such as ``The probability that a person other than the defendant, randomly selected from the population, will have this profile is 1 in 5 billion,'' whereas PGS produces results such as ``The probability of observing this evidence if the DNA comes from three unknown, unrelated contributors is 5 billion times more likely than the probability of observing this evidence if the DNA comes from the defendant and two unknown, unrelated contributors.''\footnote{The former statistic is the \textit{random match probability (RMP)}, while the latter is a \textit{likelihood ratio}. Both are statistical approaches to analyzing forensic DNA evidence~\cite{butler2021dna}.} If a public defender unfamiliar with the differences between these statistics receives a report with the latter statement, but with no explicit mention of the words ``probabilistic genotyping software,'' and doesn't know to associate the latter with PGS, they would not be able to identify that the government used PGS.

Additional difficulties our participants faced in keeping up to date with government adoption of new technologies could further exacerbate these challenges. Several participants explained that the government rarely notified them about adoption of new technologies and described how they often found out about new technologies by word-of-mouth and in individual cases. For instance, when describing how they found out about the local forensic lab's adoption of PGS, P5 felt that, \textit{``nobody notifie[d] [them] that they're using something new,''} and that they specifically found out through \textit{``an expert that [P5] use[s] for DNA cases,''} who told P5 that \textit{``she had heard that they were validating some kind of probabilistic genotyping equipment.''} When P5 directly asked the lab about their PGS use, P5 \textit{``was told that that they were beyond the validation stage, and they were using [PGS].''} In addition to learning about new technologies through word of mouth, P5 additionally described that they typically learned about new technologies \textit{``via arrest affidavits.''} Other participants expressed similar experiences finding out about new technologies through individual cases [P3, P8].

\subsubsection{Important information does not exist}\label{sec:challenge:info-dne}
Even when participants knew about government use of CFS, they often encountered difficulties assessing software reliability when information they needed about the software or software use did not exist. Our findings reveal three specific examples of this challenge: a lack of documentation about software development and use, a lack of independent empirical tests under settings representative of how the software is used, and a lack of case-specific software outputs conditioned on different theories of the case. 

Insufficient documentation practices in software development and use hindered several participants' efforts to assess software reliability [P13, P15]. For instance, P15 described a past case in which they asked for bug reports from the software company, only to find out that they did not exist. Similarly, reflecting on their past efforts to learn details about how police officers used facial recognition, P13 explained: \textit{``they don't have proper procedures in place for what they're really supposed to do when they're using [facial recognition]. They're kind of lax about the procedures and what needs to be logged, what an officer needs to keep down, what they need to maintain after they've used the software, like they only have to keep a screenshot of the first eight results. Whereas there's like 200 results that the system actually brings up. So even when we can get discovery about it, it's pretty limited and not actually tracking all the important information.''}

Participants also highlighted that insufficient information about software performance under settings representative of their use in casework could limit the utility of existing evaluation results in assessing CFS reliability in their case at hand [P4, P9]. For example, P4 described that they had not seen any studies of a widely used gunshot detection software that test the software in big cities with large buildings and loud traffic sounds that might impact the software's ability to locate sounds inferred to be gunshots. P9 raised similar concerns with PGS validation studies, describing how they felt that developers were not testing PGS with DNA mixtures of similar complexity to those often used in casework. Both of these findings echo prior work highlighting gaps between CFS test settings and casework settings~\cite{garvie2022forensic, butler2021dna, krane2022using}. Several participants additionally expressed concerns about the lack of performance evaluations conducted by independent groups, as many of the validation studies they had seen for probabilistic genotyping and gunshot detection software had been conducted by groups with financial or professional interests in promoting the use of these CFS [P13, P14].

Finally, participants sometimes wanted to know how the PGS output would change when run with different parameters, but could not make this assessment when labs only ran these software systems under one set of hypotheses --- typically the one proposed by the prosecution [P12, P15]. Since our participants often lacked access to the software, training, and materials required to re-run the software themselves (see Section~\ref{sec:challenges-accessing-info}), the set of possible outputs discussed in the case was often constrained by the lab's initial assessments. While defense could, and sometimes did, ask labs to re-run the software with different parameters, this could be infeasible if the lab was unable or unwilling to do so --- a difficulty we introduce in Section~\ref{sec:challenges-accessing-info}.

\subsubsection{Challenges getting access to existing information} \label{sec:challenges-accessing-info}Even when information did exist, actions by police departments, prosecutors, and forensic labs could still limit the extent to which defenders were able to assess reliability. Our participants expressed frustration at trade secret claims, restrictive nondisclosure agreements and slow responses to information requests, and police and forensic labs' ties to prosecutors.

Sometimes, our participant faced barriers that prevented them from getting any access to existing information. Many participants expressed their frustrations at CFS companies and users claiming that information such as descriptions of how the software works, software source code, and software executables were trade secrets and therefore could not be disclosed to the defense [P2, P4, P6, P7, P13]. Our participants' experiences echoed challenges documented and discussed in prior work on use of trade secret claims in the U.S. criminal legal system~\cite{wexler2018life, siems2022trade}.

Even when participants' efforts to access information were not entirely blocked by trade secret claims, participants found that the utility of the information they had access to could be constrained by restrictive nondisclosure agreements and delays in receiving information. When reflecting on one company's past practices, P15 remarked that the nondisclosure agreements the company used to make experts sign \textit{``were truly bonkers''} as they \textit{``implied that under a strict reading of contractual terms, an expert could look at something, discover something was wrong, and the [NDA] prohibited them from telling the court that something was wrong without the company agreeing.''} Delays in accessing information posed particular challenges in cases where defendants were in jail while they awaited the government's response to their requests for information [P7, P11, P15]. Delays were sometimes due to prosecutors incorrectly interpreting the defender's request [P15]. Other times, requests were interpreted without issue, but defenders still did not receive the information until right before trial [P7]. These delays not only extended the amount of time their clients awaited trial in jail, but also limited public defenders' ability to rigorously scrutinize the information presented to them.
    
Lastly, software users' (e.g., forensic labs and police departments) ties with prosecutors created additional roadblocks in public defenders' attempts to gather information. For example, P12 described how forensic labs ran PGS systems according to hypotheses presented to them by the prosecution and didn't test alternative hypotheses that might be favorable to the defense\footnote{As we described in Section~\ref{sec:background}, PGS takes two competing hypotheses as input and estimates the likelihood of observing the DNA evidence conditioned on each hypothesis. An example of P12's concern is a case in which the prosecution has the lab test a hypothesis that the defendant and two unknown, unrelated others contributed to the mixture, while the defense might want to test a hypothesis that the defendant, their sibling, and one unknown other contributed to the mixture.}. P12 expressed their frustration, saying: \textit{``[The forensic lab] shouldn't be trying to prove the prosecutor's hypothesis. They should be trying to see what makes the most sense scientifically.''} P15 felt similar concerns in a past case and, as a result, had asked for records of conversations between the prosecutor and the lab as a way to assess the extent to which conversations with prosecutors influenced the lab's decisions. 

To address these challenges, public defenders sometimes sought to consult with analysts at the lab and ask that they conduct additional tests with different hypotheses. However, doing so could undermine the defense, because it could reveal their strategy to the prosecution. While some participants described that they never had difficulties consulting with their local forensic labs [P16, P17], several participants felt that the forensic labs they encountered in past cases were not neutral and sided with prosecutors [P2, P3, P12, P15]. For example, P12 described how forensic labs would often notify prosecutors of defenders' requests to re-run the software and seek permission (from the prosecution) before doing so. Labs sharing this information with the prosecution can create difficulties for public defenders, since defense may want to keep their strategy a secret from the prosecution. Several participants highlighted the importance of this secrecy [P3, P4, P6, P12], which P3 succinctly illustrated: \textit{``If you're telling people kind of early on, [that] I found a problem with the DNA, the prosecutor is going to try to fix [the problem]''}. P12 similarly raised a similar concern when describing a path forward: \textit{``[T]he defense has no obligation to give the laboratory its hypothesis and they shouldn't be required to, but the laboratory on its own should consider setting up different hypotheses and running it in different ways.''}

\subsubsection{Lack of legal avenues to contest reliability} \label{sec:challenge:legal-avenues}
In addition to limiting the information public defenders were able to get, we find that decisions by law enforcement and prosecutors in how they formally described their use of CFS in court could leave defenders with no legal avenues to contest CFS reliability. We illustrate this challenge by focusing specifically on our participants' descriptions of their experiences in past cases that involved law enforcement use of facial recognition.

Several participants described past cases in which they felt that law enforcement had relied heavily on facial recognition to arrest their client while claiming they \textit{``only used [facial recognition] for investigative purposes''} [P7] (i.e., as an investigative lead) [P4, P7, P8, P10, P13]. Because law enforcement described uses of facial recognition as ``investigative'' and therefore not as trial evidence, public defenders could not file motions for admissibility hearings that would let them challenge the reliability of the software use and output, as admissibility hearings only apply to information to be introduced as evidence at trial\footnote{A \textit{motion} is ``a formal request made by any party for a desired ruling, order, or judgment''~\cite{wexlaw_motion}. An \textit{admissibility hearing} is a pre-trial hearing in which the prosecution and defense argue over whether a piece of evidence is admissible --- i.e., evidence that may be presented before the trier of fact (e.g., the jury) for them to consider in deciding the case~\cite{wexlaw_admissible_evidence}. A defense attorney who wants to contest the reliability of the prosecution's CFS evidence may bring a motion to the judge to grant an admissibility hearing for the CFS evidence. If the judge rules that the evidence is inadmissible, the CFS evidence will not be introduced at trial, meaning the jury will not see the evidence. However, admissibility hearings only apply to information introduced as evidence.}. Consequently, while public defenders felt that the software output played a significant role in identifying their client for arrest, they lacked the legal avenues to contest its reliability. P4 describes this challenge in more detail, while illustrating how police use facial recognition:

\aptLtoX[graphic=no,type=html]{\begin{quote}\textit{[Police] will get a [ranked] list of possible candidates based on the system's algorithm. [\ldots] Then an officer [\ldots] will look at this candidate list, make his own decision on which one he thinks is the most likely to be the person, [and] will then provide that to the case detective saying, `[T]his is a lead, a possible match. Not enough to make an arrest, but it's a lead. Go investigate this person.' [O]ften what they'll do next is put that photo in [\ldots] a photo array where they show six different people's photos to a witness and say, `Do you recognize any of these people?' If the witness says `Yes, I recognize that person, it's the person who committed the crime against me (or I watched commit a crime),' that would be considered an identification assuming it's the same person [that] the facial recognition came up with and then they will arrest the person based on that. [\ldots] The way the courts look at it, it doesn't really matter what came before that put that person in the photo array, as long as the witness --- as long as the procedure of the photo array itself was not unduly prejudicial.}\end{quote}}{\blockquote{\textit{[Police] will get a [ranked] list of possible candidates based on the system's algorithm. [\ldots] Then an officer [\ldots] will look at this candidate list, make his own decision on which one he thinks is the most likely to be the person, [and] will then provide that to the case detective saying, `[T]his is a lead, a possible match. Not enough to make an arrest, but it's a lead. Go investigate this person.' [O]ften what they'll do next is put that photo in [\ldots] a photo array where they show six different people's photos to a witness and say, `Do you recognize any of these people?' If the witness says `Yes, I recognize that person, it's the person who committed the crime against me (or I watched commit a crime),' that would be considered an identification assuming it's the same person [that] the facial recognition came up with and then they will arrest the person based on that. [\ldots] The way the courts look at it, it doesn't really matter what came before that put that person in the photo array, as long as the witness --- as long as the procedure of the photo array itself was not unduly prejudicial.}}}

P15 elaborated on the specific legal challenges that P4 alluded to:
\aptLtoX[graphic=no,type=html]{\begin{quote}\textit{There are certain types of motions that you only get to file if, for example, evidence is gonna be introduced in a case. And [\ldots] the facial recognition results aren't going to be introduced. What's gonna be introduced is the eyewitness identification of somebody picking your client out of a photo array. [\ldots] You don't get to challenge that earlier stage [where facial recognition was used], is how law enforcement and prosecutors tend to frame it.}\end{quote}}{\blockquote{\textit{There are certain types of motions that you only get to file if, for example, evidence is gonna be introduced in a case. And [\ldots] the facial recognition results aren't going to be introduced. What's gonna be introduced is the eyewitness identification of somebody picking your client out of a photo array. [\ldots] You don't get to challenge that earlier stage [where facial recognition was used], is how law enforcement and prosecutors tend to frame it.}}}

In response to this challenge, P4 described trying to convince the judge to extend admissibility hearings to investigative uses of facial recognition. However, when asked how courts have reacted to their arguments, P4 responded, \textit{``most courts have [\ldots] not agreed with that argument. But the reality is, most trial court judges are terrible to begin with. They're very pro-prosecution.''} P4's response suggests that judges' interpretations of the rules for admissibility hearings, and any pre-existing beliefs that shape their interpretation of the rules, further exacerbate these challenges introduced by CFS users. We further discuss challenges introduced by judges in Section~\ref{sec:challenges:noncrit-perceptions}.

\subsubsection{Withdrawn evidence and hard-to-reject plea deals} \label{sec:challenge:withdrawn-evidence}
In addition to dealing with investigative uses of software that preclude opportunities to challenge their reliability, our participants also often found themselves without opportunities to rigorously scrutinize and challenge software evidence when the prosecutor or software company removed CFS evidence from a case. This could happen when either the prosecutor or the company chose to withdraw CFS evidence, or when prosecutors offered a plea deal that the public defender felt they could not reject [P3, P7, P10, P12, P13].

Our participants often felt that when they got close to mounting a significant threat to the prosecution's software evidence (e.g., by being granted access to important information or being granted an opportunity to contest the software reliability in court), one of two things would happen: (1) the prosecution or software company would withdraw the software evidence, or (2) the prosecution would offer a plea deal that was better for the defendant. These participants' experiences with withdrawn evidence align with similar situations that have been documented in journalism~\cite{feathers2021police, jouvenal2021secret}. Since their duty to seek the best outcome for their client led them to accept and sometimes even strive for such outcomes, several participants admitted they had mixed feelings about prosecutors and software companies' decisions to withdraw evidence and prosecutors' provision of plea deals in these situations [P3, P4, P7, P10, P12, P13]. While these actions helped the defendant in the case at hand, participants highlighted how withdrawn evidence and plea deals could prevent them from gaining insights about software reliability that would have helped them contest software reliability in future clients' cases [P3, P4, P7], or also, in cases settled via plea deals, insights about software reliability that could have led to better outcomes for the defendant in the case at hand (e.g., revealing the software's unreliability to the jury results in acquittal).

\subsection{Overcoming judges and jurors' non-critical perceptions of CFS}
\label{sec:challenges:noncrit-perceptions}
In addition to navigating barriers posed by policies and practices of CFS users and developers, public defenders had to convince judges and jurors --- who were tasked with making decisions in the case --- that the software output was unreliable. However, participants highlighted how judges and jurors often held non-critical perceptions of CFS, making this a difficult task. Participants felt that these non-critical perceptions of CFS were primarily driven by judges and jurors' (1) prior beliefs about (and trust in) technology, (2) perceptions of other stakeholders (i.e., law enforcement and forensic scientists), and (3) difficulties understanding technology and statistics. 

\subsubsection{Prior beliefs about technology.} 
\label{sec:challenge:prior-beliefs-about-tech}
In describing their past interactions with judges and jurors, several participants felt that judges and jurors' prior beliefs about software tools and forensic methods prevented these groups from critically examining their reliability. For example, P13 felt that many judges and jurors perceived machines as objective and therefore reliable, explaining that \textit{``if you say to somebody that like, `Oh, a machine did this', they're like, `A machine did it. A machine doesn't have feelings. A machine makes a very reliable decision about something.' And so they're like, not really willing to question it more.''} P7 saw a similar lack of skepticism, describing their perception that judges and jurors \textit{``believe that [technology] is magic,''} i.e., accepting what the technology says without questioning its accuracy or understanding how it arrived at this decision. 

The impact of prior beliefs on perceptions of CFS is especially prevalent in how participants viewed judges' perceptions of the reliability of PGS. Several participants described how judges declared PGS evidence to be admissible (i.e., sufficiently reliable for the jury to see it) because they equated the reliability of PGS evidence with that of traditional DNA evidence. In doing so, they directly applied their beliefs about traditional DNA to reason about the reliability PGS evidence [P3, P11] despite significant differences between the reliability of traditional DNA evidence and PGS evidence\footnote{Many agree upon the reliability of traditional methods of DNA analysis for high-quality, single-source DNA samples (e.g., a large amount of minimally-degraded DNA from a single individual), but the reliability of PGS for complex DNA mixtures remains the subject of much debate~\cite{butler2021dna, lander2016forensic, matthews2019right, matthews2020trusted}.}. Consequently, participants felt that judges had failed to scrutinize the reliability of PGS evidence. Participants faced similar issues with jurors' perceptions of DNA evidence. Many participants felt that jurors were less willing to question the evidence due to what P2 described as the \textit{``CSI effect''} of DNA [P2, P7, P11, P12]. For example, in the words of P7, \textit{``the jury will say `DNA, it's DNA. What do you want? CSI taught me that DNA is the answer to all problems.'''}

\subsubsection{Perceptions of other stakeholders.} In addition to a lack of skepticism driven by prior beliefs about technology, many participants felt that judges were overly deferential to prosecutors and police officers who, along with forensic practitioners, are the predominant users of CFS [P2, P4, P6, P7, P9]. Participants reasoned that, as a result, this deference led judges to not critically examine the reliability of the CFS evidence. For example, P13 describes that \textit{``there's not many judges with the spine to push back against [these software]''} because of police departments' ability to influence judicial appointments: \textit{``When you're talking about tools used by law enforcement, if a judge here were to suddenly turn around and actually say, `Wait, there's something really wrong here. You can't use this, you shouldn't be using it', that means they're pitting themselves up against the entire [police department] and it's a political position. They are appointed judges --- they just might not get appointed again if they decide that this tool that law enforcement has claimed is necessary for them to do their job --- if they make it impossible for them to use it.''} This deference may also arise for other reasons (e.g., from judges who are former prosecutors [P9]).

\subsubsection{Difficulties understanding technology and statistics.} 
\label{sec:challenges:difficulties-understanding-tech}
Finally, even with prior beliefs of technology and perceptions of stakeholders put aside, several participants highlighted how judges and jurors' inability to understand important technical details about CFS could lead to non-critical views of software reliability. For example, when reflecting on a newer, more complex PGS tool being more difficult to challenge than an older PGS tool, P14 highlighted how the increased complexity \textit{``[made] it harder to get a judge to recognize [reliability] issue[s]''}. In reflecting on a colleague's past case, P3 explained how jurors faced difficulties understanding the public defender's arguments about the DNA evidence in the case.

Similar difficulties in understanding technology arose in our discussions with participants about judges and jurors' misunderstanding statistics such as likelihood ratios output by PGS systems and error rates. For example, when discussing jurors' understanding of likelihood ratios, participants highlighted how jurors have a hard time understanding the \textit{``technical nuance''} [P15] of these statistics and, as a result, are easily swayed by large likelihood ratios that they mistakenly associate with certainty [P9, P15]. As P15 explained, \textit{``what they're gonna say is like, `But the likelihood ratio is 800 octillion [\ldots] and that's a really big number and I have no idea what that number is, but it's a big number. And so what that sounds like is certainty.'''} P15 additionally explained their perception that jurors like binaries and consequently may have a harder time understanding likelihood ratios and error rates: \textit{``We like probabilistic things, they're more accurate but they also are harder for juries to understand because juries like binaries. They like `match, not a match.' They don't like, [\ldots] `I don't know with certainty, but it is more likely than not that it's a match.'''} Participants also raised concerns about judges and jurors' interpretations of very small numbers, referring to currently discussed error rates that were typically on the order of 0.01\% or smaller. Defenders specifically worried that judges and jurors would fail to understand the implications of small error rates like 0.001\% and not \textit{``give [it] any import''} [P11, P15].

Our participants sometimes expressed skepticism at how these existing error rates were being calculated, pointing to details they felt that CFS developers and users were omitting. For example, P9 recounted one case in which they felt that error rates for one forensic method should have accounted for inconclusive results. P15 additionally described, in the context of facial recognition, how false positives and false negatives could be defined in numerous ways, and how such definitions would be subject to debate: \textit{``The first thing you have to do is come up with a definition for what qualifies as a false positive or a false negative. And there's a lot of disagreement. [For] example, in the false negative, [\ldots] there's a lot of debate when you're calculating what those numbers are [\ldots]. Is it a false negative if the person who should have been identified is in the top ten results? Do you wind up with nine false positives? [Or] is that not a false positive problem, because in our top ten, the right person was there? [In] defining what counts, there's gonna be a lot of debate.''} However, P15 revealed how this desire to discuss the nuances of these calculations and calculate more detailed, rigorous measurements of performance could be hampered by judges and jurors having difficulty understanding statistics: \textit{``[Clear definitions and justifications for them] could be really, really helpful. But judges [and jurors] are not statisticians and so they get a little lost in [technical debates such as] what is the difference between a false positive and false negative? And why would we be more concerned about one than the other?''} 

\subsection{Gathering CFS expertise from scientists, technologists, and other lawyers}
\label{sec:challenge:experts}
Reflecting on the challenges they face in convincing judges and jurors to critically examine CFS reliability, all of the public defenders we spoke to emphasized that having an expert witnesses who could explain these concepts to judges and jurors was crucial to their success. Many public defenders also highlighted the importance of working with outside experts when gathering information and developing their arguments, and several participants additionally described how their offices were building in-house expertise through in-house technologists and attorneys specializing in forensics. However, our participants faced various barriers to acquiring these different forms of expertise, ranging from not knowing to work with an expert, to challenges in finding and funding expertise.

\subsubsection{Not seeking experts}
\label{sec:challenge:not-seeking-experts} Our interviews revealed that public defenders may forgo working with experts related to a CFS tool that was used in a case. On the one hand, public defenders may not seek such an expert because they do not know that a CFS is being used in the case at hand, an issue we described in Section~\ref{sec:challenge:identifying-sw-use}. On the other hand, they may know about the CFS tool used but may not reach out to an expert [P2, P3, P7]. P3 described how this may happen due to the attorney believing that they do not need an expert. P2 and P7 described how this may also happen when a defense attorney sees DNA, thinks the prosecution has a strong case against their client (the ``CSI effect of DNA'' that we described in Section~\ref{sec:challenge:prior-beliefs-about-tech}), and then decides to accept a plea deal or forgo scrutinize the PGS tool. 

\subsubsection{Limited expert availability and willingness to work with defense}\label{sec:challenge:limited-expert-availability} Even in instances when public defenders sought outside expertise, they sometimes faced difficulties finding independent experts who both had deep working knowledge relevant to the specific issues at hand and were willing to work with the defense.

Several participants described difficulties in finding independent experts since many people with the necessary expertise work for CFS companies or government, who both present conflicts of interest [P4, P10, P12]. CFS employees' financial interests are in securing partnerships with buyers, i.e., law enforcement and forensic labs, and in convincing others that their software is reliable. P10 expressed their concerns with government actors such as analysts from forensic labs: \textit{``And, you know, the worry with somebody who works for the government is that they're paid by the government, [\ldots] and the government is also trying to prosecute the client. And so there's a little bit of an inherent conflict of interest there, where they might be afraid to lose their jobs, for agreeing with you that the algorithm is biased or whatever.''}

In addition to concerns about the independence of experts, payment timelines and expert availability posed additional challenges to finding experts. For example, in P6's jurisdiction, defenders \textit{``don't have any ability to pay experts up front, so they have to find people who are willing to work and wait\ldots for the [court] to ultimately pay them at the end of the process.''} P6 expressed frustration, saying: \textit{``that's tough, just because [\ldots] people don't want to work on spec.''} Several participants detailed the challenges they or their colleagues faced in finding experts with time to work with them [P6, P8, P12, P14]. As P14 described, it is \textit{``challenging finding experts who have really deep working knowledge and the few that do are so piled on by the rest of the country [are] so overworked.''}

While our findings do not represent experiences in all jurisdictions, participants revealed how courts --- which often provide public defenders with funding for experts --- can further complicate the process of obtaining outside expertise. P6 described how state-level changes in court procedures made it easier for public defenders to get funds for experts \textit{``by passing a law that says that [defenders] can ask for those funds ex-parte which means outside of the presence of the other party. [So we can make] a showing to the court that we need to have a hearing without the prosecutors there. And then we can tell the judge why we need expert funds, which keeps us from having to tell the prosecutors you know stuff about our case that we shouldn't have to tell them or stuff about our preparation.''} Jurisdictions without these guarantees may face additional difficulties obtaining funding for experts through courts.

\subsubsection{Difficulties building institutional knowledge via in-house experts, attorney-analyst liaisons, learning from other public defenders} \label{sec:challenge:difficulties-building-institutional-knowledge}Participants also highlighted three other approaches to gathering CFS expertise for criminal cases, beyond working with outside experts: hiring in-house technologists and analysts, creating roles for public defenders in the office who specialize in acting as liaisons between other public defenders and either outside or in-house experts, and sourcing strategies from public defenders working in other offices and jurisdictions [P4, P5, P8].

When asked about how they perceived the difference between not having in-house experts and having them, P8 emphasized that \textit{``it's night and day''}, since having in-house experts streamlined public defenders' tasks such as getting \textit{``almost instant feedback''} [P8] from experts on potential courtroom strategies and helping the expert prepare for trial testimony and cross-examination. However, despite these perceived benefits of having in-house experts, these participants were often referring to situations in which they perceived the public defender office in question to be relatively well-funded. In most other situations, insufficient funding continues to hinder public defense offices' ability to fund these additional positions.

Some participants additionally highlighted the importance of another source of CFS expertise: public defenders who act as liaisons between other attorneys and experts (both in-house and outside). P4 explained the importance of these attorney-analyst liaisons, or \textit{``tech-savvy attorneys''} [P4] through illustrating a disconnect P4 had observed between analysts and attorneys: \textit{``They both seemed to be talking different languages and not understanding each other [\ldots] the analysts [\ldots] understood the technology, and how it worked, but didn't necessarily understand --- to the extent that was needed --- exactly how it's used in court, [\ldots] and what legal issues may arise from different things that they were doing. [\ldots] The attorneys on the other hand [\ldots] --- very few were anywhere near the level of the analysts' tech savviness. And so there was a lot of not understanding what the analysts are saying and how it can be used in our case.''} Summarizing these challenges, P4 concluded that \textit{``The attorney is talking legal talk and the analyst is talking tech talk, and kind of talking over each other without really finding how those two things meet and how it would be useful and helpful for clients.''} The attorney-analyst liaison, as P4 argued, helped bridge this disconnect. Several of our participants described similar responsibilities as being part of their current roles [P5, P8, P9, P10, P13, P14]. However, not all of these participants carried out these responsibilities as paid positions. For instance, P5 described \textit{``[coordinating] a lot of the forensics in the office''} as \textit{``not part of [their] job,''}, and that they were \textit{``trying to create that position''}. P5's situation, contrasted with other participants' descriptions of their liaison-work as being a formal part of their job as attorneys, suggests that insufficient funding and extreme caseloads may additionally pose difficulties in building internal CFS expertise through analyst-attorney liaisons.

Lastly, many of our participants also often referred to their experiences learning from public defenders in other offices and other jurisdictions. P3 explained that they had \textit{``talked to people around the country who have done [PGS] cases''} to learn about different strategies other attorneys have adopted, and how courts and jurors reacted to those arguments. But P3 highlighted that sourcing strategies from defenders working in other states faced limitations, as legal mechanisms, judges, and demographics of jury pools of the specific jurisdiction could all impact the success of a given strategy. In the words of P3, \textit{``Until you stand up and go to court and operate under our [state's] law, [\ldots] what's going to happen [in practice] is the big unknown.''}
\section{Discussion}
\label{sec:discussion}
In this study, we found that public defenders faced a variety of challenges assessing CFS reliability and contesting CFS outputs in court. In this section, we summarize the main findings, document how each relates to prior research, and discuss implications for efforts to make performance evaluations effective for public defenders seeking to assess and contest CFS reliability in the U.S. criminal legal system.

\subsection{Critically examine how factors outside the design and communication of performance evaluations can constrain opportunities to scrutinize and contest reliability.}

We find that \textbf{factors beyond the design and communication of the performance evaluation could prevent public defenders from raising concerns about CFS reliability in the first place} (Sections~\ref{sec:challenge:identifying-sw-use},~\ref{sec:challenge:legal-avenues},~\ref{sec:challenge:withdrawn-evidence}). These findings align with discussions in recent work in human-AI interaction and responsible AI revealing how laws and policies, organizational pressures, and limited knowledge and expertise can preclude impacted communities, frontline workers, and advocates from scrutinizing and contesting algorithm reliability (e.g.,~\cite{kawakami2022improving, kuo2023understanding, alfrink2023contestable}). 

Efforts to make performance evaluations effective for downstream stakeholders should not only aim to understand how to design and communicate evaluations so they address these individuals’ needs and goals, but also understand how the contexts in which these individuals operate might preclude them from making use of this information. To guide this future work, we build on the aforementioned prior work and discuss how (i) narrow definitions of in-scope technology use, (ii) conflicting goals, and (iii) limited information about other aspects of AI design, development, and use can preclude opportunities to scrutinize and contest algorithm reliability.

\subsubsection{Understand how definitions of in-scope technology uses, and interpretations of these definitions, can constrain opportunities to assess and contest algorithmically-driven decisions} 

We find that \textbf{the scope of technology uses considered in laws and policies, and decision-makers' interpretations of these laws and policies, can constrain efforts to assess and contest CFS reliability}. Specifically, we see legal rules, and judges' interpretations of them, treat investigative uses of CFS as unimportant and uncontestable compared to evidentiary uses of CFS (Sections~\ref{sec:challenge:legal-avenues},~\ref{sec:challenges:noncrit-perceptions}), despite the significant influence that investigative uses of facial recognition can have in law enforcement decisions about who to arrest~\cite{garvie2022forensic}. Prior work has similarly discussed the impacts of how laws and policies define in-scope technologies and uses of technology~\cite{richardson2021defining, krafft2020defining, wright2024null, lawrence2023bureaucratic}, and how various stakeholders interpret these definitions~\cite{krafft2020defining, metcalf2023taking}. For instance,~\citet{krafft2020defining} find that policymakers tend to define ``AI'' by comparing systems to human thinking and behavior, while AI researchers tend towards definitions that emphasize technical functionality. The authors highlight that, as a result, policies may overemphasize concerns about future technologies at the expense of urgent issues with existing technologies~\cite{krafft2020defining}.  Within the context of CFS in the U.S. criminal legal system, we recommend that policies defining in-scope technologies based on their relation to ``evidence'' (e.g.,~\cite{takano2021}) carefully consider technologies whose outputs may strongly influence case outcomes but are not formally introduced as evidence. More broadly, future research at the intersection of policy design and human-AI interaction design should carefully examine definitions of in-scope technologies and technology uses, various stakeholders' interpretations of these definitions, how these definitions may occlude important concerns from decision-making, and the impact of such omissions on downstream efforts to scrutinize and contest automated decision systems.

\subsubsection{Understand how advocates' and frontline workers' desires to assess and contest reliability may conflict with their other goals} 
We find that \textbf{a public defender's duty and desire to do what is best for their clients could, at times, come into conflict with their desire to scrutinize the reliability of CFS systems used against their clients}. For instance, some participants felt mixed feelings when CFS users and developers removed CFS evidence from a case: accepting a plea deal or continuing the case without the incriminatory evidence might have been best for their client in the case at hand, but doing so also prevented defenders from gaining insights about CFS reliability that might have helped them contest CFS reliability in future cases (Section~\ref{sec:challenge:withdrawn-evidence}). This tension between doing what is best for their current client and creating opportunities to raise reliability concerns in future clients' cases echoes trade-offs documented in recent HCI work studying the experiences of frontline workers in child welfare and housing contexts. Collectively, these studies find that frontline workers' efforts to question or contest the reliability of algorithmic outputs may come at the cost of less time spent on other cases~\cite{kuo2023understanding}, and accusations by administrators of disagreeing with the AI system too often~\cite{kawakami2022improving, saxena2021framework}. In the context of public defenders' scrutiny of CFS in the U.S. criminal legal system, we argue that the current lack of independent validation of CFS~\cite{canellas2021defending, butler2021dna} places the burden of assessing CFS reliability on the shoulders of public defenders and is a key contributor to the tensions our participants felt. Consequently, we echo prior work in calling for policies that require rigorous, independent evaluations of CFS, \textit{before} they are used in criminal casework. More broadly, our findings and insights from prior work illustrate how the needs and goals of frontline workers may sometimes come into conflict with their desires to scrutinize algorithm reliability. In light of these challenges, we recommend that future work seeking to design systems that support contestation by those advocating on behalf of decision subjects balance these tensions in ways that enable advocates to contest decisions in ways that neither undermine the case at hand, nor limit their ability to scrutinize these systems to inform future contestation efforts.

\subsubsection{Understand how policies for documenting information about the development, evaluation, and use of automated decision systems can constrain advocates' contestation efforts}
\label{sec:disc:barrier-limited-info}
While our participants experienced a variety of challenges in getting access to existing information about CFS systems and outputs (Section~\ref{sec:challenges-accessing-info}), we additionally find that \textbf{public defenders and those who work with them sometimes sought out information --- such as bug reports, system performance under settings similar to those in the case at hand, and build information --- only to find that the information did not exist} (Section~\ref{sec:challenge:info-dne}). These insights empirically ground ongoing discussions at the intersection of AI accountability and public policy arguing for better documentation of AI uses and harms~\cite{metcalf2023taking, alfrink2023contestable, costanza2022audits}. Future work designing human-AI interactions can support these calls for policy interventions by providing further insights into the information needs of downstream stakeholders and how documentation practices might fall short of these needs in practice. For example, recent work studying human-AI collaboration in the child welfare system found that social workers desired information beyond the outputs of the automated decision systems they work with --- such as factors the algorithm considered in producing its output --- when deciding whether to trust a given output~\cite{kawakami2022care, kawakami2022improving}. These findings, along with the methods the authors used to understand workers' needs, provide a starting point that we believe can support further understanding of overcoming these information barriers.\\

\noindent Even when participants did obtain opportunities to assess and contest reliability, the challenges they encountered or anticipated in using performance information and communicating their concerns to judges and jurors reveal important considerations for designing and communicating performance evaluations of CFS and automated decision systems to support contestation, which we discuss in the Sections~\ref{sec:disc:eval-design} and~\ref{sec:disc:eval-communication}, respectively.

\subsection{Contextualize design of performance evaluations in real world applications}\label{sec:disc:eval-design}

In exploring our participants' experiences scrutinizing CFS, we find that public defenders \textbf{perceived crucial gaps between test inputs and the real-world use cases they encountered in their day-to-day work.} For instance, participants highlighted that tests for gunshot detection software and PGS should take into account the local geography and population demographics, respectively, of their specific jurisdictions (Section~\ref{sec:challenge:info-dne}). For CFS systems that produced a likelihood ratio dependent on input hypotheses, participants recalled how they wanted to run the system with alternative hypotheses that they felt would better capture the details of the case (Sections~\ref{sec:challenge:info-dne},~\ref{sec:challenges-accessing-info}). \textbf{Participants also expressed desires for CFS evaluations to use a broader range of metrics beyond accuracy and false positive error rates to assess system limitations.} Example metrics that participants proposed included error rates for determinations that a piece of evidence was inconclusive and ranges for the possible values that false positive software outputs could take on (Section~\ref{sec:challenges:difficulties-understanding-tech}). Together, these findings echo gaps between evaluation design and real-world use contexts highlighted in prior work (e.g.,~\cite{yang2023designing, hutchinson2022critical, raji2021ai, suresh2023kaleidoscope}), and, importantly, demonstrate the valuable insights that public defenders develop through their everyday encounters with CFS in the U.S. criminal legal system, further motivating growing efforts in HCI to engage downstream stakeholders in designing performance evaluations of AI systems (e.g.,~\cite{suresh2023kaleidoscope, shen2021everyday, devos2022toward, lam2022end}). 

In this section, we build on this prior literature by discussing two categories of implications for future work that seeks to support public defender engagement in the design of CFS performance evaluations: (i) uncover and address barriers to engaging frontline workers in evaluation design and (ii) supporting skill-sharing and cross-disciplinary collaboration. Whenever relevant, we additionally draw on and contribute to recent work discussing approaches and challenges to participatory design of AI systems (e.g.,~\cite{delgado2023participatory, birhane2022power}).

\subsubsection{Uncover and address barriers to engaging frontline workers in evaluation design} Recent work in user-engaged auditing and contextualized evaluation highlights the crucial insights that groups with experiential and domain knowledge (e.g., everyday users and domain experts) can bring to the design of performance evaluations~\cite{suresh2023kaleidoscope, deng2023understanding}. While public defenders possess crucial experiential expertise that can ensure more rigorous testing of CFS reliability, our participants illustrated how insufficient documentation and disclosure practices may result in public defenders seeing only a subset of instances, or a subset of details about instances, in which the government uses CFS (Sections~\ref{sec:challenge:identifying-sw-use},~\ref{sec:challenge:info-dne}). In other words, \textbf{our work highlights how automated decision systems may still be used in contexts outside the purview of stakeholders with experiential expertise, consequently limiting opportunities for their full engagement in participatory design of performance evaluations}. Future work should seek to understand when and how policies and practices in using automated decision systems may obscure certain uses of these systems, as this opacity may constrain frontline workers' ability to fully engage in evaluation design.

\subsubsection{Support cross-disciplinary collaboration and skill-sharing} In addition to illustrating barriers that could hinder public defense participation in the design of contextualized evaluations, our work also highlights \textbf{a key aspect of public defenders’ needs and practices that future work should take into account when designing evaluations with and evaluation tools for public defenders: the collaboration and skill-sharing practices they leverage in scrutinizing and contesting CFS.} Recent work introducing tools and frameworks for user-driven evaluations offers promising directions for future work seeking to support public defenders in conducting their own evaluations of CFS according to their own definitions of important test cases and software performance~\cite{suresh2023kaleidoscope, abebe2022adversarial}. We build on this work by distinguishing between two types of collaboration and skill-sharing our participants partake in. 

Some of this collaboration and skill-sharing occurs among defense attorneys (Section~\ref{sec:challenge:difficulties-building-institutional-knowledge}), who likely share similar skillsets and legal training (e.g., public defenders sharing strategies and insights with other defenders) and is consequently similar to those documented in~\citet{kawakami2022improving}, where authors found that social workers in the child welfare system collaborated with each other to understand the inner workings of the risk assessment algorithm they used. However, we also find that \textbf{public defenders heavily rely on cross-disciplinary collaboration with CFS experts who are familiar with technology but may be less familiar with law} (Section~\ref{sec:challenge:difficulties-building-institutional-knowledge}). These findings highlight a need to better understand how tools and frameworks for user-driven evaluations can support collaboration and skill-sharing amongst stakeholders with different disciplinary backgrounds and different types of expertise. For future work exploring these topics, we additionally highlight an important parallel between the skill-sharing and cross-disciplinary collaboration discussed here and an aim of participatory methods to spread knowledge of technical experts and of communities with experiential expertise~\cite{birhane2022power}. We suggest that future work closely engages with these discussions of methods for facilitating and empowering such knowledge transfer while necessarily grappling with participants' incentives and the various costs (e.g., time, energy) associated with deep, sustained participatory engagement.

\subsection{Understand real-world factors that shape interpretation of performance results}\label{sec:disc:eval-communication}
We find that public defenders additionally faced challenges when communicating their concerns with CFS reliability to judges and jurors. These challenges reveal three research areas that we recommend future work explores when seeking to design and communicate performance evaluations such that they are effective for downstream stakeholders who aim to scrutinize and contest the reliability of automated decision systems.

\subsubsection{Understand how deliberation of performance information influences perceptions of reliability} Prior work in HCI has studied how laypeople understand performance information and has shown that providing people with information about model accuracy and uncertainty can help lay decision-makers appropriately calibrate trust. However, this prior work typically investigates perceptions of reliability by presenting information (e.g., the accuracy rate or some measure of uncertainty) to an individual software user, who then decides whether to trust the algorithm’s output (e.g., ~\cite{yin2019understanding, prabhudesai2023understanding, yang2023subjective}). We find that \textbf{deliberation and contestation of such performance outputs may play a crucial role in mediating real-world perceptions of algorithmic performance}. 

For example, participants highlighted how defense and prosecution would inevitably argue over what the appropriate definition of an error rate should be (Section~\ref{sec:challenges:difficulties-understanding-tech}). In these scenarios where the two sides disagree over how to define the error rate, judges and jurors would hear not just one error rate, but contrasting proposals for what the appropriate error rate should be. Consequently, the jury's perception of the reliability of the given output is not mediated by a single error rate, but instead, by multiple proposals for what an appropriate error rate should be. For instance, recalling P15's comments on this scenario in the context of facial recognition, the jury may hear that one side argue that a false positive occurs when a non-matching individual is included in the top n results, while the other side argues that a false positive occurs only when a non-matching individual is the top result returned from the facial recognition system. These two definitions and the jury's observation of the process of deliberation between the arguing parties would, together, shape the jury's perception of the software output's reliability.

Our findings complement two key motivations behind contextualized evaluations: the performance of an AI system cannot be captured in a single metric~\cite{barocas2021designing}, and different stakeholders carry different notions of what acceptable model performance looks like~\cite{suresh2023kaleidoscope}. This insight suggests that deliberating choices behind performance evaluations extend beyond the U.S. criminal legal system. Consequently, we recommend that future work studying CFS and other automated decision systems further explores this multi-faceted communication process and its effects on people's perceptions of performance. Initial steps that may help align this work with real-world contexts are empirical studies that explore how people currently present and discuss error rates in different downstream contexts.

\subsubsection{Understand how presentation of performance information influences perceptions of reliability} 
In addition to finding that decision-makers receive multiple interpretations of algorithm performance, we find that even the presentation of a single perspective may complicate how people perceive reliability. We specifically find that \textbf{the complexity of performance metrics may shape people's perceptions of reliability}. Public defenders expressed a desire to communicate the nuances of evaluation choices and performance information to contest the reliability of CFS outputs in their cases, but also felt the need to simplify concepts for judges and jurors (Section~\ref{sec:challenges:difficulties-understanding-tech}). Future work should explore how the presentation of performance information, and decision-makers' perceptions of its complexity, can shape their perceptions of reliability.

Our study also highlights how \textbf{other information presented to the decision-maker may also influence their perceptions of reliability.} In the context of CFS, we found that, in addition to error rates, likelihood ratios may sway or influence decision-makers' perceptions of output reliability. As~\citet{garrett2020error} highlight in their study of jurors' perceptions of forensic method reliability, likelihood ratios will often be presented alongside error rates. This presentation of other information may also arise in response to findings in recent human-AI collaboration literature revealing that human decision-makers seek additional information beyond accuracy when assessing the reliability of algorithmic outputs~\cite{cai2019hello}. Consequently, it is crucial to understand how this additional information influences how people interpret performance information. Assessing the impacts of other information is especially important when decision-makers may incorrectly use this other information to understand reliability. When discussing PGS used to analyze DNA evidence, our participants highlighted how jurors perceived likelihood ratios output by PGS as measures of certainty, despite prior work highlighting that the magnitude of a likelihood ratio itself does not communicate certainty~\cite{lund2017likelihood}.

\subsubsection{Understanding how prior beliefs and knowledge influences perceptions of reliability}
Beyond processes of deliberation and individual presentations of performance information, \textbf{our participants believed that decision-makers’ prior beliefs of technology and technology users also mediated perceptions of performance}. For example, our participants described how judges and jurors’ beliefs of technology as magic and the strength of forensic techniques, combined with difficulties these decision-makers faced in understanding technical concepts and statistical nuances, may have led these decision-makers to adopt non-critical perceptions of the reliability of these software outputs (Section~\ref{sec:challenges:noncrit-perceptions}). Prior work has empirically studied how these prior beliefs about the technology at hand, familiarity with AI, and knowledge about the task domain impact people's perceptions of system reliability (e.g.,~\cite{garrett2020error, zhang2020effect, ehsan2021explainable, goyal2023else}). For instance, experimental studies have looked at how participant identities such as their familiarity with technology or ML shapes their ability to appropriately calibrate trust~\cite{zhang2020effect}. 

However, building on this existing literature, we find that \textbf{additional priors, specifically existing perceptions of technology users and related disciplines such as forensic practitioners and forensic methods may influence perceptions of reliability.} To further illustrate this point, we compare our findings with findings from prior work documenting how judges do not use data analytics or risk assessment tools~\cite{christin2017algorithms}. While judges' non-critical perceptions of CFS may seem to conflict with these observations of judges' aversion to data-driven algorithms, we explain how these findings in~\cite{christin2017algorithms} further support our findings. In discussing judges' aversion to risk assessment tools,~\citet{christin2017algorithms} describe a general resistance to innovation in criminal justice, along with legal professionals' aversion to using tools ``built by a company they know nothing about.'' CFS, while introducing new computational methods to forensic analysis, benefits from judges and attorneys' existing perceptions of forensics as a tried and true discipline in the U.S. criminal legal system (Section~\ref{sec:challenges:noncrit-perceptions}). These findings, taken together, highlight opportunities to further investigate how these additional dimensions of individuals' prior beliefs and knowledge influence their perceptions of the reliability of algorithmic outputs, both in the case of CFS in the U.S. criminal legal system and in other domains.
\section{Conclusion}
In this paper, we have focused on the real-world experiences of public defenders contesting automated decision systems in one high-stakes setting --- the U.S. criminal legal system --- on behalf of those accused of crimes. We specifically study the challenges public defenders face when assessing and contesting the reliability of computational forensic software that the government increasingly relies on to convict and incarcerate. Our findings suggest that efforts to leverage performance evaluations to contest algorithmic decisions may be constrained by a wide range of technical, social, and institutional barriers. Future work should center the technical, social, and institutional contexts in which contestation occurs, to better position performance evaluations to support contestability in practice --- an aim that becomes increasingly urgent and crucial as automated decision systems are increasingly deployed in high-stakes settings.

\begin{acks}
We would like to thank our interview participants for their time and valuable input that shaped this research. We also thank our anonymous reviewers for their dedication and thorough feedback that substantially improved this paper. This material is based upon work supported by the National Science Foundation Graduate Research Fellowship Program under Grant No. DGE 2146752, by NSF award IIS-2129008, and by the UC Berkeley AI Policy Hub, an academic initiative advancing interdisciplinary research to anticipate and address AI policy opportunities. Any opinions, findings, and conclusions or recommendations expressed in this material are those of the authors and do not necessarily reflect the views of the National Science Foundation.
\end{acks}

\bibliographystyle{ACM-Reference-Format}
\bibliography{main}


\begin{thebibliography}{95}


\ifx \showCODEN    \undefined \def \showCODEN     #1{\unskip}     \fi
\ifx \showDOI      \undefined \def \showDOI       #1{#1}\fi
\ifx \showISBNx    \undefined \def \showISBNx     #1{\unskip}     \fi
\ifx \showISBNxiii \undefined \def \showISBNxiii  #1{\unskip}     \fi
\ifx \showISSN     \undefined \def \showISSN      #1{\unskip}     \fi
\ifx \showLCCN     \undefined \def \showLCCN      #1{\unskip}     \fi
\ifx \shownote     \undefined \def \shownote      #1{#1}          \fi
\ifx \showarticletitle \undefined \def \showarticletitle #1{#1}   \fi
\ifx \showURL      \undefined \def \showURL       {\relax}        \fi
\providecommand\bibfield[2]{#2}
\providecommand\bibinfo[2]{#2}
\providecommand\natexlab[1]{#1}
\providecommand\showeprint[2][]{arXiv:#2}

\bibitem[eff(2017)]%
        {eff_alpr}
 \bibinfo{year}{2017}\natexlab{}.
\newblock \bibinfo{title}{Automated License Plate Readers (ALPRs)}.
\newblock \bibinfo{howpublished}{\url{https://www.eff.org/pages/automated-license-plate-readers-alpr}}.
\newblock
\newblock
\shownote{Accessed: 2022-09-13}.


\bibitem[wex(2021)]%
        {wexlaw_admissible_evidence}
 \bibinfo{year}{2021}\natexlab{}.
\newblock \bibinfo{title}{Admissible evidence}.
\newblock \bibinfo{howpublished}{\emph{Wex, by the Legal Information Institute at Cornell University}. Accessed September 13, 2023}.
\newblock
\urldef\tempurl%
\url{"https://www.law.cornell.edu/wex/admissible_evidence)."}
\showURL{%
\tempurl}


\bibitem[str(2022)]%
        {strmixUse}
 \bibinfo{year}{2022}\natexlab{}.
\newblock \bibinfo{title}{STRmix Use in Certified U.S. Forensic Labs Continues to Grow}.
\newblock \bibinfo{howpublished}{\url{https://www.strmix.com/news/strmix-use-in-certified-u-s-forensic-labs-continues-to-grow/}}.
\newblock
\newblock
\shownote{Accessed: 2023-09-13}.


\bibitem[acl(2023)]%
        {aclu_fr_scrutiny}
 \bibinfo{year}{2023}\natexlab{}.
\newblock \bibinfo{title}{After Third Wrongful Arrest, ACLU Slams Detroit Police Department for Continuing to Use Faulty Facial Recognition Technology}.
\newblock \bibinfo{howpublished}{\url{https://www.aclu.org/press-releases/after-third-wrongful-arrest-aclu-slams-detroit-police-department-for-continuing-to-use-faulty-facial-recognition-technology}}.
\newblock
\newblock
\shownote{Accessed: 2022-09-13}.


\bibitem[wex(2023)]%
        {wexlaw_motion}
 \bibinfo{year}{2023}\natexlab{}.
\newblock \bibinfo{title}{Motion}.
\newblock \bibinfo{howpublished}{\emph{Wex, by the Legal Information Institute at Cornell University}. Accessed September 13, 2023}.
\newblock
\urldef\tempurl%
\url{"https://www.law.cornell.edu/wex/motion"}
\showURL{%
\tempurl}


\bibitem[Abebe et~al\mbox{.}(2022)]%
        {abebe2022adversarial}
\bibfield{author}{\bibinfo{person}{Rediet Abebe}, \bibinfo{person}{Moritz Hardt}, \bibinfo{person}{Angela Jin}, \bibinfo{person}{John Miller}, \bibinfo{person}{Ludwig Schmidt}, {and} \bibinfo{person}{Rebecca Wexler}.} \bibinfo{year}{2022}\natexlab{}.
\newblock \showarticletitle{Adversarial scrutiny of evidentiary statistical software}. In \bibinfo{booktitle}{\emph{Proceedings of the 2022 ACM Conference on Fairness, Accountability, and Transparency}}. \bibinfo{pages}{1733--1746}.
\newblock


\bibitem[Alfrink et~al\mbox{.}(2023)]%
        {alfrink2023contestable}
\bibfield{author}{\bibinfo{person}{Kars Alfrink}, \bibinfo{person}{Ianus Keller}, \bibinfo{person}{Neelke Doorn}, {and} \bibinfo{person}{Gerd Kortuem}.} \bibinfo{year}{2023}\natexlab{}.
\newblock \showarticletitle{Contestable Camera Cars: A Speculative Design Exploration of Public AI That Is Open and Responsive to Dispute}. In \bibinfo{booktitle}{\emph{Proceedings of the 2023 CHI Conference on Human Factors in Computing Systems}}. \bibinfo{pages}{1--16}.
\newblock


\bibitem[Alfrink et~al\mbox{.}(2022)]%
        {alfrink2022contestable}
\bibfield{author}{\bibinfo{person}{Kars Alfrink}, \bibinfo{person}{Ianus Keller}, \bibinfo{person}{Gerd Kortuem}, {and} \bibinfo{person}{Neelke Doorn}.} \bibinfo{year}{2022}\natexlab{}.
\newblock \showarticletitle{Contestable AI by Design: Towards a Framework}.
\newblock \bibinfo{journal}{\emph{Minds and Machines}} (\bibinfo{year}{2022}), \bibinfo{pages}{1--27}.
\newblock


\bibitem[Alkhatib(2021)]%
        {alkhatib2021live}
\bibfield{author}{\bibinfo{person}{Ali Alkhatib}.} \bibinfo{year}{2021}\natexlab{}.
\newblock \showarticletitle{To live in their utopia: Why algorithmic systems create absurd outcomes}. In \bibinfo{booktitle}{\emph{Proceedings of the 2021 CHI conference on human factors in computing systems}}. \bibinfo{pages}{1--9}.
\newblock


\bibitem[Almada(2019)]%
        {almada2019human}
\bibfield{author}{\bibinfo{person}{Marco Almada}.} \bibinfo{year}{2019}\natexlab{}.
\newblock \showarticletitle{Human intervention in automated decision-making: Toward the construction of contestable systems}. In \bibinfo{booktitle}{\emph{Proceedings of the Seventeenth International Conference on Artificial Intelligence and Law}}. \bibinfo{pages}{2--11}.
\newblock


\bibitem[Bansal et~al\mbox{.}(2019)]%
        {bansal2019beyond}
\bibfield{author}{\bibinfo{person}{Gagan Bansal}, \bibinfo{person}{Besmira Nushi}, \bibinfo{person}{Ece Kamar}, \bibinfo{person}{Walter~S Lasecki}, \bibinfo{person}{Daniel~S Weld}, {and} \bibinfo{person}{Eric Horvitz}.} \bibinfo{year}{2019}\natexlab{}.
\newblock \showarticletitle{Beyond accuracy: The role of mental models in human-AI team performance}. In \bibinfo{booktitle}{\emph{Proceedings of the AAAI conference on human computation and crowdsourcing}}, Vol.~\bibinfo{volume}{7}. \bibinfo{pages}{2--11}.
\newblock


\bibitem[Barocas et~al\mbox{.}(2021)]%
        {barocas2021designing}
\bibfield{author}{\bibinfo{person}{Solon Barocas}, \bibinfo{person}{Anhong Guo}, \bibinfo{person}{Ece Kamar}, \bibinfo{person}{Jacquelyn Krones}, \bibinfo{person}{Meredith~Ringel Morris}, \bibinfo{person}{Jennifer~Wortman Vaughan}, \bibinfo{person}{W~Duncan Wadsworth}, {and} \bibinfo{person}{Hanna Wallach}.} \bibinfo{year}{2021}\natexlab{}.
\newblock \showarticletitle{Designing disaggregated evaluations of ai systems: Choices, considerations, and tradeoffs}. In \bibinfo{booktitle}{\emph{Proceedings of the 2021 AAAI/ACM Conference on AI, Ethics, and Society}}. \bibinfo{pages}{368--378}.
\newblock


\bibitem[Beyer and Holtzblatt(1999)]%
        {beyer1999contextual}
\bibfield{author}{\bibinfo{person}{Hugh Beyer} {and} \bibinfo{person}{Karen Holtzblatt}.} \bibinfo{year}{1999}\natexlab{}.
\newblock \showarticletitle{Contextual design}.
\newblock \bibinfo{journal}{\emph{interactions}} \bibinfo{volume}{6}, \bibinfo{number}{1} (\bibinfo{year}{1999}), \bibinfo{pages}{32--42}.
\newblock


\bibitem[Birhane et~al\mbox{.}(2022)]%
        {birhane2022power}
\bibfield{author}{\bibinfo{person}{Abeba Birhane}, \bibinfo{person}{William Isaac}, \bibinfo{person}{Vinodkumar Prabhakaran}, \bibinfo{person}{Mark Diaz}, \bibinfo{person}{Madeleine~Clare Elish}, \bibinfo{person}{Iason Gabriel}, {and} \bibinfo{person}{Shakir Mohamed}.} \bibinfo{year}{2022}\natexlab{}.
\newblock \showarticletitle{Power to the people? opportunities and challenges for participatory AI}.
\newblock \bibinfo{journal}{\emph{Equity and Access in Algorithms, Mechanisms, and Optimization}} (\bibinfo{year}{2022}), \bibinfo{pages}{1--8}.
\newblock


\bibitem[Buolamwini and Gebru(2018)]%
        {buolamwini2018gender}
\bibfield{author}{\bibinfo{person}{Joy Buolamwini} {and} \bibinfo{person}{Timnit Gebru}.} \bibinfo{year}{2018}\natexlab{}.
\newblock \showarticletitle{Gender shades: Intersectional accuracy disparities in commercial gender classification}. In \bibinfo{booktitle}{\emph{Conference on fairness, accountability and transparency}}. PMLR, \bibinfo{pages}{77--91}.
\newblock


\bibitem[Burke et~al\mbox{.}(2022)]%
        {shotspotterUse}
\bibfield{author}{\bibinfo{person}{Garance Burke}, \bibinfo{person}{Martha Mendoza}, \bibinfo{person}{Juliet Linderman}, {and} \bibinfo{person}{Michael Tarm}.} \bibinfo{year}{2022}\natexlab{}.
\newblock \bibinfo{title}{How AI-powered tech landed man in jail with scant evidence}.
\newblock \bibinfo{howpublished}{\url{https://apnews.com/article/artificial-intelligence-algorithm-technology-police-crime-7e3345485aa668c97606d4b54f9b6220}}.
\newblock
\newblock
\shownote{Accessed: 2022-09-13}.


\bibitem[Butler et~al\mbox{.}(2021)]%
        {butler2021dna}
\bibfield{author}{\bibinfo{person}{John~M Butler}, \bibinfo{person}{Hari Iyer}, \bibinfo{person}{Rich Press}, \bibinfo{person}{Melissa Taylor}, \bibinfo{person}{Peter~M Vallone}, {and} \bibinfo{person}{Sheila Willis}.} \bibinfo{year}{2021}\natexlab{}.
\newblock \showarticletitle{DNA Mixture Interpretation: A NIST Scientific Foundation Review}.
\newblock \bibinfo{journal}{\emph{National Institutes of Standards and Technology}} (\bibinfo{year}{2021}).
\newblock


\bibitem[Cabrera et~al\mbox{.}(2023)]%
        {cabrera2023zeno}
\bibfield{author}{\bibinfo{person}{{\'A}ngel~Alexander Cabrera}, \bibinfo{person}{Erica Fu}, \bibinfo{person}{Donald Bertucci}, \bibinfo{person}{Kenneth Holstein}, \bibinfo{person}{Ameet Talwalkar}, \bibinfo{person}{Jason~I Hong}, {and} \bibinfo{person}{Adam Perer}.} \bibinfo{year}{2023}\natexlab{}.
\newblock \showarticletitle{Zeno: An interactive framework for behavioral evaluation of machine learning}. In \bibinfo{booktitle}{\emph{Proceedings of the 2023 CHI Conference on Human Factors in Computing Systems}}. \bibinfo{pages}{1--14}.
\newblock


\bibitem[Cai et~al\mbox{.}(2019)]%
        {cai2019hello}
\bibfield{author}{\bibinfo{person}{Carrie~J Cai}, \bibinfo{person}{Samantha Winter}, \bibinfo{person}{David Steiner}, \bibinfo{person}{Lauren Wilcox}, {and} \bibinfo{person}{Michael Terry}.} \bibinfo{year}{2019}\natexlab{}.
\newblock \showarticletitle{" Hello AI": uncovering the onboarding needs of medical practitioners for human-AI collaborative decision-making}.
\newblock \bibinfo{journal}{\emph{Proceedings of the ACM on Human-computer Interaction}} \bibinfo{volume}{3}, \bibinfo{number}{CSCW} (\bibinfo{year}{2019}), \bibinfo{pages}{1--24}.
\newblock


\bibitem[Canellas(2021)]%
        {canellas2021defending}
\bibfield{author}{\bibinfo{person}{Marc Canellas}.} \bibinfo{year}{2021}\natexlab{}.
\newblock \showarticletitle{Defending IEEE Software Standards in Federal Criminal Court}.
\newblock \bibinfo{journal}{\emph{Computer}} \bibinfo{volume}{54}, \bibinfo{number}{6} (\bibinfo{year}{2021}), \bibinfo{pages}{14--23}.
\newblock


\bibitem[Charmaz(2006)]%
        {charmaz2006constructing}
\bibfield{author}{\bibinfo{person}{Kathy Charmaz}.} \bibinfo{year}{2006}\natexlab{}.
\newblock \bibinfo{booktitle}{\emph{Constructing grounded theory: A practical guide through qualitative analysis}}.
\newblock \bibinfo{publisher}{sage}.
\newblock


\bibitem[Christin(2017)]%
        {christin2017algorithms}
\bibfield{author}{\bibinfo{person}{Ang{\`e}le Christin}.} \bibinfo{year}{2017}\natexlab{}.
\newblock \showarticletitle{Algorithms in practice: Comparing web journalism and criminal justice}.
\newblock \bibinfo{journal}{\emph{Big Data \& Society}} \bibinfo{volume}{4}, \bibinfo{number}{2} (\bibinfo{year}{2017}), \bibinfo{pages}{2053951717718855}.
\newblock


\bibitem[Coble and Bright(2019)]%
        {coble2019probabilistic}
\bibfield{author}{\bibinfo{person}{Michael~D Coble} {and} \bibinfo{person}{Jo-Anne Bright}.} \bibinfo{year}{2019}\natexlab{}.
\newblock \showarticletitle{Probabilistic genotyping software: an overview}.
\newblock \bibinfo{journal}{\emph{Forensic Science International: Genetics}}  \bibinfo{volume}{38} (\bibinfo{year}{2019}), \bibinfo{pages}{219--224}.
\newblock


\bibitem[Congress.gov(2021)]%
        {takano2021}
\bibfield{author}{\bibinfo{person}{Library of~Congress Congress.gov}.} \bibinfo{year}{2021}\natexlab{}.
\newblock \bibinfo{booktitle}{\emph{H.R.2438 - 117th Congress (2021-2022): Justice in Forensic Algorithms Act of 2021.}}
\newblock
\urldef\tempurl%
\url{https://www.congress.gov/bill/117th-congress/house-bill/2438}
\showURL{%
Retrieved February 23, 2023 from \tempurl}


\bibitem[Costanza-Chock et~al\mbox{.}(2022)]%
        {costanza2022audits}
\bibfield{author}{\bibinfo{person}{Sasha Costanza-Chock}, \bibinfo{person}{Inioluwa~Deborah Raji}, {and} \bibinfo{person}{Joy Buolamwini}.} \bibinfo{year}{2022}\natexlab{}.
\newblock \showarticletitle{Who Audits the Auditors? Recommendations from a field scan of the algorithmic auditing ecosystem}. In \bibinfo{booktitle}{\emph{Proceedings of the 2022 ACM Conference on Fairness, Accountability, and Transparency}}. \bibinfo{pages}{1571--1583}.
\newblock


\bibitem[Delgado et~al\mbox{.}(2023)]%
        {delgado2023participatory}
\bibfield{author}{\bibinfo{person}{Fernando Delgado}, \bibinfo{person}{Stephen Yang}, \bibinfo{person}{Michael Madaio}, {and} \bibinfo{person}{Qian Yang}.} \bibinfo{year}{2023}\natexlab{}.
\newblock \showarticletitle{The Participatory Turn in AI Design: Theoretical Foundations and the Current State of Practice}. In \bibinfo{booktitle}{\emph{Proceedings of the 3rd ACM Conference on Equity and Access in Algorithms, Mechanisms, and Optimization}}. \bibinfo{pages}{1--23}.
\newblock


\bibitem[Deng et~al\mbox{.}(2023)]%
        {deng2023understanding}
\bibfield{author}{\bibinfo{person}{Wesley~Hanwen Deng}, \bibinfo{person}{Boyuan Guo}, \bibinfo{person}{Alicia Devrio}, \bibinfo{person}{Hong Shen}, \bibinfo{person}{Motahhare Eslami}, {and} \bibinfo{person}{Kenneth Holstein}.} \bibinfo{year}{2023}\natexlab{}.
\newblock \showarticletitle{Understanding Practices, Challenges, and Opportunities for User-Engaged Algorithm Auditing in Industry Practice}. In \bibinfo{booktitle}{\emph{Proceedings of the 2023 CHI Conference on Human Factors in Computing Systems}}. \bibinfo{pages}{1--18}.
\newblock


\bibitem[DeVos et~al\mbox{.}(2022)]%
        {devos2022toward}
\bibfield{author}{\bibinfo{person}{Alicia DeVos}, \bibinfo{person}{Aditi Dhabalia}, \bibinfo{person}{Hong Shen}, \bibinfo{person}{Kenneth Holstein}, {and} \bibinfo{person}{Motahhare Eslami}.} \bibinfo{year}{2022}\natexlab{}.
\newblock \showarticletitle{Toward User-Driven Algorithm Auditing: Investigating users’ strategies for uncovering harmful algorithmic behavior}. In \bibinfo{booktitle}{\emph{Proceedings of the 2022 CHI Conference on Human Factors in Computing Systems}}. \bibinfo{pages}{1--19}.
\newblock


\bibitem[Ehsan et~al\mbox{.}(2021)]%
        {ehsan2021explainable}
\bibfield{author}{\bibinfo{person}{Upol Ehsan}, \bibinfo{person}{Samir Passi}, \bibinfo{person}{Q~Vera Liao}, \bibinfo{person}{Larry Chan}, \bibinfo{person}{I Lee}, \bibinfo{person}{Michael Muller}, \bibinfo{person}{Mark~O Riedl}, {et~al\mbox{.}}} \bibinfo{year}{2021}\natexlab{}.
\newblock \showarticletitle{The who in explainable ai: How ai background shapes perceptions of ai explanations}.
\newblock \bibinfo{journal}{\emph{arXiv preprint arXiv:2107.13509}} (\bibinfo{year}{2021}).
\newblock


\bibitem[Farole and Langton(2010)]%
        {farole2010county}
\bibfield{author}{\bibinfo{person}{Donald~J Farole} {and} \bibinfo{person}{Lynn Langton}.} \bibinfo{year}{2010}\natexlab{}.
\newblock \bibinfo{booktitle}{\emph{County-based and local public defender offices, 2007}}.
\newblock \bibinfo{publisher}{US Department of Justice, Office of Justice Programs, Bureau of Justice~…}.
\newblock


\bibitem[Feathers(2021)]%
        {feathers2021police}
\bibfield{author}{\bibinfo{person}{Todd Feathers}.} \bibinfo{year}{2021}\natexlab{}.
\newblock \showarticletitle{Police are telling ShotSpotter to alter evidence from gunshot-detecting AI}.
\newblock \bibinfo{journal}{\emph{Retrieved November}}  \bibinfo{volume}{23} (\bibinfo{year}{2021}), \bibinfo{pages}{2022}.
\newblock


\bibitem[Furst(2019)]%
        {furst2019fair}
\bibfield{author}{\bibinfo{person}{Bryan Furst}.} \bibinfo{year}{2019}\natexlab{}.
\newblock \showarticletitle{A Fair Fight}.
\newblock \bibinfo{journal}{\emph{Brennan Center for Justice, September}}  \bibinfo{volume}{9} (\bibinfo{year}{2019}).
\newblock


\bibitem[Garrett et~al\mbox{.}(2020)]%
        {garrett2020error}
\bibfield{author}{\bibinfo{person}{Brandon~L Garrett}, \bibinfo{person}{William~E Crozier}, {and} \bibinfo{person}{Rebecca Grady}.} \bibinfo{year}{2020}\natexlab{}.
\newblock \showarticletitle{Error rates, likelihood ratios, and jury evaluation of forensic evidence}.
\newblock \bibinfo{journal}{\emph{Journal of Forensic Sciences}} \bibinfo{volume}{65}, \bibinfo{number}{4} (\bibinfo{year}{2020}), \bibinfo{pages}{1199--1209}.
\newblock


\bibitem[Garvie(2016)]%
        {garvie2016perpetual}
\bibfield{author}{\bibinfo{person}{Clare Garvie}.} \bibinfo{year}{2016}\natexlab{}.
\newblock \bibinfo{booktitle}{\emph{The perpetual line-up: Unregulated police face recognition in America}}.
\newblock \bibinfo{publisher}{Georgetown Law, Center on Privacy \& Technology}.
\newblock


\bibitem[Garvie(2022)]%
        {garvie2022forensic}
\bibfield{author}{\bibinfo{person}{Clare Garvie}.} \bibinfo{year}{2022}\natexlab{}.
\newblock \showarticletitle{A forensic without the science: face recognition in US criminal investigations}.
\newblock \bibinfo{journal}{\emph{Center on Privacy \& Technology at Georgetown Law}}  \bibinfo{volume}{6} (\bibinfo{year}{2022}).
\newblock


\bibitem[Gaub et~al\mbox{.}(2019)]%
        {gaub2019understanding}
\bibfield{author}{\bibinfo{person}{Janne~E Gaub}, \bibinfo{person}{Carolyn Naoroz}, {and} \bibinfo{person}{Aili Malm}.} \bibinfo{year}{2019}\natexlab{}.
\newblock \showarticletitle{Understanding the impact of police body-worn cameras on Virginia public defenders}.
\newblock \bibinfo{journal}{\emph{Retrieved March}}  \bibinfo{volume}{2} (\bibinfo{year}{2019}), \bibinfo{pages}{2022}.
\newblock


\bibitem[Goyal et~al\mbox{.}(2023)]%
        {goyal2023else}
\bibfield{author}{\bibinfo{person}{Navita Goyal}, \bibinfo{person}{Eleftheria Briakou}, \bibinfo{person}{Amanda Liu}, \bibinfo{person}{Connor Baumler}, \bibinfo{person}{Claire Bonial}, \bibinfo{person}{Jeffrey Micher}, \bibinfo{person}{Clare~R Voss}, \bibinfo{person}{Marine Carpuat}, {and} \bibinfo{person}{Hal Daum{\'e}~III}.} \bibinfo{year}{2023}\natexlab{}.
\newblock \showarticletitle{What Else Do I Need to Know? The Effect of Background Information on Users' Reliance on AI Systems}.
\newblock \bibinfo{journal}{\emph{arXiv preprint arXiv:2305.14331}} (\bibinfo{year}{2023}).
\newblock


\bibitem[Harlow(2001)]%
        {harlow2001defense}
\bibfield{author}{\bibinfo{person}{Caroline~Wolf Harlow}.} \bibinfo{year}{2001}\natexlab{}.
\newblock \bibinfo{booktitle}{\emph{Defense counsel in criminal cases}}.
\newblock \bibinfo{publisher}{US Department of Justice, Office of Justice Programs, Bureau of Justice~…}.
\newblock


\bibitem[Hildebrandt(2019)]%
        {hildebrandt2019privacy}
\bibfield{author}{\bibinfo{person}{Mireille Hildebrandt}.} \bibinfo{year}{2019}\natexlab{}.
\newblock \showarticletitle{Privacy as protection of the incomputable self: From agnostic to agonistic machine learning}.
\newblock \bibinfo{journal}{\emph{Theoretical Inquiries in Law}} \bibinfo{volume}{20}, \bibinfo{number}{1} (\bibinfo{year}{2019}), \bibinfo{pages}{83--121}.
\newblock


\bibitem[Hill(2020)]%
        {facialrecogUse}
\bibfield{author}{\bibinfo{person}{Kashmir Hill}.} \bibinfo{year}{2020}\natexlab{}.
\newblock \bibinfo{title}{Wrongfully Accused by an Algorithm}.
\newblock \bibinfo{howpublished}{\url{https://www.nytimes.com/2020/06/24/technology/facial-recognition-arrest.html}}.
\newblock
\newblock
\shownote{Accessed: 2022-09-13}.


\bibitem[Hirsch et~al\mbox{.}(2017)]%
        {hirsch2017designing}
\bibfield{author}{\bibinfo{person}{Tad Hirsch}, \bibinfo{person}{Kritzia Merced}, \bibinfo{person}{Shrikanth Narayanan}, \bibinfo{person}{Zac~E Imel}, {and} \bibinfo{person}{David~C Atkins}.} \bibinfo{year}{2017}\natexlab{}.
\newblock \showarticletitle{Designing contestability: Interaction design, machine learning, and mental health}. In \bibinfo{booktitle}{\emph{Proceedings of the 2017 Conference on Designing Interactive Systems}}. \bibinfo{pages}{95--99}.
\newblock


\bibitem[Hutchinson et~al\mbox{.}(2022)]%
        {hutchinson2022critical}
\bibfield{author}{\bibinfo{person}{Ben Hutchinson}, \bibinfo{person}{Christina Greer}, \bibinfo{person}{Katherine Heller}, \bibinfo{person}{Negar Rostamzadeh}, {and} \bibinfo{person}{Vinodkumar Prabhakaran}.} \bibinfo{year}{2022}\natexlab{}.
\newblock \showarticletitle{Critical Evaluation Gaps in Machine Learning Practice}.
\newblock  (\bibinfo{year}{2022}).
\newblock


\bibitem[Joh(2017)]%
        {joh2017undue}
\bibfield{author}{\bibinfo{person}{Elizabeth~E Joh}.} \bibinfo{year}{2017}\natexlab{}.
\newblock \showarticletitle{The undue influence of surveillance technology companies in policing}.
\newblock \bibinfo{journal}{\emph{NYUL Rev. Online}}  \bibinfo{volume}{92} (\bibinfo{year}{2017}), \bibinfo{pages}{19}.
\newblock


\bibitem[Jouvenal(2021)]%
        {jouvenal2021secret}
\bibfield{author}{\bibinfo{person}{J Jouvenal}.} \bibinfo{year}{2021}\natexlab{}.
\newblock \showarticletitle{A secret algorithm is transforming DNA evidence. This defendant could be the first to scrutinize it}.
\newblock \bibinfo{journal}{\emph{Washington Post, July}}  \bibinfo{volume}{13} (\bibinfo{year}{2021}), \bibinfo{pages}{2021}.
\newblock


\bibitem[Kaminski and Urban(2021)]%
        {kaminski2021right}
\bibfield{author}{\bibinfo{person}{Margot~E Kaminski} {and} \bibinfo{person}{Jennifer~M Urban}.} \bibinfo{year}{2021}\natexlab{}.
\newblock \showarticletitle{The right to contest AI}.
\newblock \bibinfo{journal}{\emph{Columbia Law Review}} \bibinfo{volume}{121}, \bibinfo{number}{7} (\bibinfo{year}{2021}), \bibinfo{pages}{1957--2048}.
\newblock


\bibitem[Kawakami et~al\mbox{.}(2022a)]%
        {kawakami2022improving}
\bibfield{author}{\bibinfo{person}{Anna Kawakami}, \bibinfo{person}{Venkatesh Sivaraman}, \bibinfo{person}{Hao-Fei Cheng}, \bibinfo{person}{Logan Stapleton}, \bibinfo{person}{Yanghuidi Cheng}, \bibinfo{person}{Diana Qing}, \bibinfo{person}{Adam Perer}, \bibinfo{person}{Zhiwei~Steven Wu}, \bibinfo{person}{Haiyi Zhu}, {and} \bibinfo{person}{Kenneth Holstein}.} \bibinfo{year}{2022}\natexlab{a}.
\newblock \showarticletitle{Improving human-AI partnerships in child welfare: understanding worker practices, challenges, and desires for algorithmic decision support}. In \bibinfo{booktitle}{\emph{Proceedings of the 2022 CHI Conference on Human Factors in Computing Systems}}. \bibinfo{pages}{1--18}.
\newblock


\bibitem[Kawakami et~al\mbox{.}(2022b)]%
        {kawakami2022care}
\bibfield{author}{\bibinfo{person}{Anna Kawakami}, \bibinfo{person}{Venkatesh Sivaraman}, \bibinfo{person}{Logan Stapleton}, \bibinfo{person}{Hao-Fei Cheng}, \bibinfo{person}{Adam Perer}, \bibinfo{person}{Zhiwei~Steven Wu}, \bibinfo{person}{Haiyi Zhu}, {and} \bibinfo{person}{Kenneth Holstein}.} \bibinfo{year}{2022}\natexlab{b}.
\newblock \showarticletitle{“Why Do I Care What’s Similar?” Probing Challenges in AI-Assisted Child Welfare Decision-Making through Worker-AI Interface Design Concepts}. In \bibinfo{booktitle}{\emph{Designing Interactive Systems Conference}}. \bibinfo{pages}{454--470}.
\newblock


\bibitem[Kirchner(2017a)]%
        {fstUse}
\bibfield{author}{\bibinfo{person}{Lauren Kirchner}.} \bibinfo{year}{2017}\natexlab{a}.
\newblock \bibinfo{title}{Thousands of Criminal Cases in New York Relied on Disputed DNA Testing Techniques}.
\newblock \bibinfo{howpublished}{\url{https://www.propublica.org/article/thousands-of-criminal-cases-in-new-york-relied-on-disputed-dna-testing-techniques}}.
\newblock
\newblock
\shownote{Accessed: 2022-09-13}.


\bibitem[Kirchner(2017b)]%
        {kirchner2017thousands}
\bibfield{author}{\bibinfo{person}{Lauren Kirchner}.} \bibinfo{year}{2017}\natexlab{b}.
\newblock \showarticletitle{Thousands of Criminal Cases in New York Relied on Disputed DNA Testing Techniques}.
\newblock \bibinfo{journal}{\emph{ProPublica (4 Sept. 2017)}} (\bibinfo{year}{2017}).
\newblock


\bibitem[Kluttz et~al\mbox{.}(2022)]%
        {kluttz2022shaping}
\bibfield{author}{\bibinfo{person}{Daniel~N Kluttz}, \bibinfo{person}{Nitin Kohli}, {and} \bibinfo{person}{Deirdre~K Mulligan}.} \bibinfo{year}{2022}\natexlab{}.
\newblock \showarticletitle{Shaping our tools: Contestability as a means to promote responsible algorithmic decision making in the professions}.
\newblock \bibinfo{journal}{\emph{Ethics of Data and Analytics. Auerbach Publications}} (\bibinfo{year}{2022}), \bibinfo{pages}{420--428}.
\newblock


\bibitem[Krafft et~al\mbox{.}(2020)]%
        {krafft2020defining}
\bibfield{author}{\bibinfo{person}{PM Krafft}, \bibinfo{person}{Meg Young}, \bibinfo{person}{Michael Katell}, \bibinfo{person}{Karen Huang}, {and} \bibinfo{person}{Ghislain Bugingo}.} \bibinfo{year}{2020}\natexlab{}.
\newblock \showarticletitle{Defining AI in policy versus practice}. In \bibinfo{booktitle}{\emph{Proceedings of the AAAI/ACM Conference on AI, Ethics, and Society}}. \bibinfo{pages}{72--78}.
\newblock


\bibitem[Krane and Philpott(2022)]%
        {krane2022using}
\bibfield{author}{\bibinfo{person}{Dan~E Krane} {and} \bibinfo{person}{M~Katherine Philpott}.} \bibinfo{year}{2022}\natexlab{}.
\newblock \showarticletitle{Using Laboratory Validation to Identify and Establish Limits to the Reliability of Probabilistic Genotyping Systems}.
\newblock In \bibinfo{booktitle}{\emph{Handbook of DNA Profiling}}. \bibinfo{publisher}{Springer}, \bibinfo{pages}{297--319}.
\newblock


\bibitem[Kuo et~al\mbox{.}(2023)]%
        {kuo2023understanding}
\bibfield{author}{\bibinfo{person}{Tzu-Sheng Kuo}, \bibinfo{person}{Hong Shen}, \bibinfo{person}{Jisoo Geum}, \bibinfo{person}{Nev Jones}, \bibinfo{person}{Jason~I Hong}, \bibinfo{person}{Haiyi Zhu}, {and} \bibinfo{person}{Kenneth Holstein}.} \bibinfo{year}{2023}\natexlab{}.
\newblock \showarticletitle{Understanding Frontline Workers’ and Unhoused Individuals’ Perspectives on AI Used in Homeless Services}. In \bibinfo{booktitle}{\emph{Proceedings of the 2023 CHI Conference on Human Factors in Computing Systems}}. \bibinfo{pages}{1--17}.
\newblock


\bibitem[Lai and Tan(2019)]%
        {lai2019human}
\bibfield{author}{\bibinfo{person}{Vivian Lai} {and} \bibinfo{person}{Chenhao Tan}.} \bibinfo{year}{2019}\natexlab{}.
\newblock \showarticletitle{On human predictions with explanations and predictions of machine learning models: A case study on deception detection}. In \bibinfo{booktitle}{\emph{Proceedings of the conference on fairness, accountability, and transparency}}. \bibinfo{pages}{29--38}.
\newblock


\bibitem[Lam et~al\mbox{.}(2022)]%
        {lam2022end}
\bibfield{author}{\bibinfo{person}{Michelle~S Lam}, \bibinfo{person}{Mitchell~L Gordon}, \bibinfo{person}{Dana{\"e} Metaxa}, \bibinfo{person}{Jeffrey~T Hancock}, \bibinfo{person}{James~A Landay}, {and} \bibinfo{person}{Michael~S Bernstein}.} \bibinfo{year}{2022}\natexlab{}.
\newblock \showarticletitle{End-User Audits: A System Empowering Communities to Lead Large-Scale Investigations of Harmful Algorithmic Behavior}.
\newblock \bibinfo{journal}{\emph{Proceedings of the ACM on Human-Computer Interaction}} \bibinfo{volume}{6}, \bibinfo{number}{CSCW2} (\bibinfo{year}{2022}), \bibinfo{pages}{1--34}.
\newblock


\bibitem[Lawrence et~al\mbox{.}(2023)]%
        {lawrence2023bureaucratic}
\bibfield{author}{\bibinfo{person}{Christie Lawrence}, \bibinfo{person}{Isaac Cui}, {and} \bibinfo{person}{Daniel Ho}.} \bibinfo{year}{2023}\natexlab{}.
\newblock \showarticletitle{The bureaucratic challenge to AI governance: An empirical assessment of implementation at US federal agencies}. In \bibinfo{booktitle}{\emph{Proceedings of the 2023 AAAI/ACM Conference on AI, Ethics, and Society}}. \bibinfo{pages}{606--652}.
\newblock


\bibitem[Liao and Xiao(2023)]%
        {liao2023rethinking}
\bibfield{author}{\bibinfo{person}{Q~Vera Liao} {and} \bibinfo{person}{Ziang Xiao}.} \bibinfo{year}{2023}\natexlab{}.
\newblock \showarticletitle{Rethinking Model Evaluation as Narrowing the Socio-Technical Gap}.
\newblock \bibinfo{journal}{\emph{arXiv preprint arXiv:2306.03100}} (\bibinfo{year}{2023}).
\newblock


\bibitem[Lund and Iyer(2017)]%
        {lund2017likelihood}
\bibfield{author}{\bibinfo{person}{Steven~P Lund} {and} \bibinfo{person}{Hari Iyer}.} \bibinfo{year}{2017}\natexlab{}.
\newblock \showarticletitle{Likelihood ratio as weight of forensic evidence: a closer look}.
\newblock \bibinfo{journal}{\emph{Journal of Research of the National Institute of Standards and Technology}}  \bibinfo{volume}{122} (\bibinfo{year}{2017}), \bibinfo{pages}{1}.
\newblock


\bibitem[Lynch({[n.\,d.]})]%
        {lynch}
\bibfield{author}{\bibinfo{person}{Katie Lynch}.} \bibinfo{year}{[n.\,d.]}\natexlab{}.
\newblock \bibinfo{title}{Know the facts}.
\newblock
\newblock
\urldef\tempurl%
\url{http://www.beingatticusfinch.com/know-the-facts}
\showURL{%
\tempurl}


\bibitem[Lyons et~al\mbox{.}(2022)]%
        {lyons2022whats}
\bibfield{author}{\bibinfo{person}{Henrietta Lyons}, \bibinfo{person}{Senuri Wijenayake}, \bibinfo{person}{Tim Miller}, {and} \bibinfo{person}{Eduardo Velloso}.} \bibinfo{year}{2022}\natexlab{}.
\newblock \showarticletitle{What’s the appeal? Perceptions of review processes for algorithmic decisions}. In \bibinfo{booktitle}{\emph{Proceedings of the 2022 CHI Conference on Human Factors in Computing Systems}}. \bibinfo{pages}{1--15}.
\newblock


\bibitem[Matthews et~al\mbox{.}(2019)]%
        {matthews2019right}
\bibfield{author}{\bibinfo{person}{Jeanna Matthews}, \bibinfo{person}{Marzieh Babaeianjelodar}, \bibinfo{person}{Stephen Lorenz}, \bibinfo{person}{Abigail Matthews}, \bibinfo{person}{Mariama Njie}, \bibinfo{person}{Nathaniel Adams}, \bibinfo{person}{Dan Krane}, \bibinfo{person}{Jessica Goldthwaite}, {and} \bibinfo{person}{Clinton Hughes}.} \bibinfo{year}{2019}\natexlab{}.
\newblock \showarticletitle{The right to confront your accusers: Opening the black box of forensic DNA software}. In \bibinfo{booktitle}{\emph{Proceedings of the 2019 AAAI/ACM Conference on AI, Ethics, and Society}}. \bibinfo{pages}{321--327}.
\newblock


\bibitem[Matthews et~al\mbox{.}(2020)]%
        {matthews2020trusted}
\bibfield{author}{\bibinfo{person}{Jeanna~Neefe Matthews}, \bibinfo{person}{Graham Northup}, \bibinfo{person}{Isabella Grasso}, \bibinfo{person}{Stephen Lorenz}, \bibinfo{person}{Marzieh Babaeianjelodar}, \bibinfo{person}{Hunter Bashaw}, \bibinfo{person}{Sumona Mondal}, \bibinfo{person}{Abigail Matthews}, \bibinfo{person}{Mariama Njie}, {and} \bibinfo{person}{Jessica Goldthwaite}.} \bibinfo{year}{2020}\natexlab{}.
\newblock \showarticletitle{When trusted black boxes don't agree: Incentivizing iterative improvement and accountability in critical software systems}. In \bibinfo{booktitle}{\emph{Proceedings of the AAAI/ACM Conference on AI, Ethics, and Society}}. \bibinfo{pages}{102--108}.
\newblock


\bibitem[Metcalf et~al\mbox{.}(2023)]%
        {metcalf2023taking}
\bibfield{author}{\bibinfo{person}{Jacob Metcalf}, \bibinfo{person}{Ranjit Singh}, \bibinfo{person}{Emanuel Moss}, \bibinfo{person}{Emnet Tafesse}, {and} \bibinfo{person}{Elizabeth~Anne Watkins}.} \bibinfo{year}{2023}\natexlab{}.
\newblock \showarticletitle{Taking Algorithms to Courts: A Relational Approach to Algorithmic Accountability}. In \bibinfo{booktitle}{\emph{Proceedings of the 2023 ACM Conference on Fairness, Accountability, and Transparency}}. \bibinfo{pages}{1450--1462}.
\newblock


\bibitem[Mitchell et~al\mbox{.}(2019)]%
        {mitchell2019model}
\bibfield{author}{\bibinfo{person}{Margaret Mitchell}, \bibinfo{person}{Simone Wu}, \bibinfo{person}{Andrew Zaldivar}, \bibinfo{person}{Parker Barnes}, \bibinfo{person}{Lucy Vasserman}, \bibinfo{person}{Ben Hutchinson}, \bibinfo{person}{Elena Spitzer}, \bibinfo{person}{Inioluwa~Deborah Raji}, {and} \bibinfo{person}{Timnit Gebru}.} \bibinfo{year}{2019}\natexlab{}.
\newblock \showarticletitle{Model cards for model reporting}. In \bibinfo{booktitle}{\emph{Proceedings of the conference on fairness, accountability, and transparency}}. \bibinfo{pages}{220--229}.
\newblock


\bibitem[Moses et~al\mbox{.}(2011)]%
        {moses2011automated}
\bibfield{author}{\bibinfo{person}{Kenneth~R Moses}, \bibinfo{person}{P Higgins}, \bibinfo{person}{M McCabe}, \bibinfo{person}{S Prabhakar}, {and} \bibinfo{person}{S Swann}.} \bibinfo{year}{2011}\natexlab{}.
\newblock \showarticletitle{Automated fingerprint identification system (AFIS)}.
\newblock \bibinfo{journal}{\emph{The fingerprint sourcebook}}  \bibinfo{volume}{1} (\bibinfo{year}{2011}), \bibinfo{pages}{6--1}.
\newblock


\bibitem[Patton et~al\mbox{.}(2020)]%
        {patton2020contextual}
\bibfield{author}{\bibinfo{person}{Desmond~U Patton}, \bibinfo{person}{William~R Frey}, \bibinfo{person}{Kyle~A McGregor}, \bibinfo{person}{Fei-Tzin Lee}, \bibinfo{person}{Kathleen McKeown}, {and} \bibinfo{person}{Emanuel Moss}.} \bibinfo{year}{2020}\natexlab{}.
\newblock \showarticletitle{Contextual analysis of social media: The promise and challenge of eliciting context in social media posts with natural language processing}. In \bibinfo{booktitle}{\emph{Proceedings of the AAAI/ACM Conference on AI, Ethics, and Society}}. \bibinfo{pages}{337--342}.
\newblock


\bibitem[{PCAST (President's Council of Advisors on Science and Technology)}(2016)]%
        {lander2016forensic}
\bibfield{author}{\bibinfo{person}{{PCAST (President's Council of Advisors on Science and Technology)}}.} \bibinfo{year}{2016}\natexlab{}.
\newblock \showarticletitle{Forensic Science in Criminal Courts: Ensuring Scientific Validation of Feature-Comparison Methods}.
\newblock \bibinfo{journal}{\emph{President’s Council of Advisors on Science and Technology}} (\bibinfo{year}{2016}).
\newblock


\bibitem[Petraco et~al\mbox{.}(2012)]%
        {petraco2012application}
\bibfield{author}{\bibinfo{person}{Nicholas~DK Petraco}, \bibinfo{person}{Helen Chan}, \bibinfo{person}{P De~Forest}, \bibinfo{person}{D Crim}, \bibinfo{person}{Peter Diaczuk}, \bibinfo{person}{Carol Gambino}, {et~al\mbox{.}}} \bibinfo{year}{2012}\natexlab{}.
\newblock \showarticletitle{Application of machine learning to toolmarks: statistically based methods for impression pattern comparisons}.
\newblock \bibinfo{journal}{\emph{National Institute of Justice}} (\bibinfo{year}{2012}).
\newblock


\bibitem[Pierre et~al\mbox{.}(2021)]%
        {pierre2021getting}
\bibfield{author}{\bibinfo{person}{Jennifer Pierre}, \bibinfo{person}{Roderic Crooks}, \bibinfo{person}{Morgan Currie}, \bibinfo{person}{Britt Paris}, {and} \bibinfo{person}{Irene Pasquetto}.} \bibinfo{year}{2021}\natexlab{}.
\newblock \showarticletitle{Getting Ourselves Together: Data-centered participatory design research \& epistemic burden}. In \bibinfo{booktitle}{\emph{Proceedings of the 2021 CHI Conference on Human Factors in Computing Systems}}. \bibinfo{pages}{1--11}.
\newblock


\bibitem[Ploug and Holm(2020)]%
        {ploug2020four}
\bibfield{author}{\bibinfo{person}{Thomas Ploug} {and} \bibinfo{person}{S{\o}ren Holm}.} \bibinfo{year}{2020}\natexlab{}.
\newblock \showarticletitle{The four dimensions of contestable AI diagnostics-A patient-centric approach to explainable AI}.
\newblock \bibinfo{journal}{\emph{Artificial Intelligence in Medicine}}  \bibinfo{volume}{107} (\bibinfo{year}{2020}), \bibinfo{pages}{101901}.
\newblock


\bibitem[Prabhudesai et~al\mbox{.}(2023)]%
        {prabhudesai2023understanding}
\bibfield{author}{\bibinfo{person}{Snehal Prabhudesai}, \bibinfo{person}{Leyao Yang}, \bibinfo{person}{Sumit Asthana}, \bibinfo{person}{Xun Huan}, \bibinfo{person}{Q~Vera Liao}, {and} \bibinfo{person}{Nikola Banovic}.} \bibinfo{year}{2023}\natexlab{}.
\newblock \showarticletitle{Understanding Uncertainty: How Lay Decision-makers Perceive and Interpret Uncertainty in Human-AI Decision Making}. In \bibinfo{booktitle}{\emph{Proceedings of the 28th International Conference on Intelligent User Interfaces}}. \bibinfo{pages}{379--396}.
\newblock


\bibitem[Raji et~al\mbox{.}(2021)]%
        {raji2021ai}
\bibfield{author}{\bibinfo{person}{Inioluwa~Deborah Raji}, \bibinfo{person}{Emily~M Bender}, \bibinfo{person}{Amandalynne Paullada}, \bibinfo{person}{Emily Denton}, {and} \bibinfo{person}{Alex Hanna}.} \bibinfo{year}{2021}\natexlab{}.
\newblock \showarticletitle{AI and the everything in the whole wide world benchmark}.
\newblock \bibinfo{journal}{\emph{arXiv preprint arXiv:2111.15366}} (\bibinfo{year}{2021}).
\newblock


\bibitem[Richardson(2021)]%
        {richardson2021defining}
\bibfield{author}{\bibinfo{person}{Rashida Richardson}.} \bibinfo{year}{2021}\natexlab{}.
\newblock \showarticletitle{Defining and demystifying automated decision systems}.
\newblock \bibinfo{journal}{\emph{Md. L. Rev.}}  \bibinfo{volume}{81} (\bibinfo{year}{2021}), \bibinfo{pages}{785}.
\newblock


\bibitem[Sarra(2020)]%
        {sarra2020put}
\bibfield{author}{\bibinfo{person}{Claudio Sarra}.} \bibinfo{year}{2020}\natexlab{}.
\newblock \showarticletitle{Put dialectics into the machine: protection against automatic-decision-making through a deeper understanding of contestability by design}.
\newblock \bibinfo{journal}{\emph{Global Jurist}} \bibinfo{volume}{20}, \bibinfo{number}{3} (\bibinfo{year}{2020}), \bibinfo{pages}{20200003}.
\newblock


\bibitem[Saxena et~al\mbox{.}(2021)]%
        {saxena2021framework}
\bibfield{author}{\bibinfo{person}{Devansh Saxena}, \bibinfo{person}{Karla Badillo-Urquiola}, \bibinfo{person}{Pamela~J Wisniewski}, {and} \bibinfo{person}{Shion Guha}.} \bibinfo{year}{2021}\natexlab{}.
\newblock \showarticletitle{A framework of high-stakes algorithmic decision-making for the public sector developed through a case study of child-welfare}.
\newblock \bibinfo{journal}{\emph{Proceedings of the ACM on Human-Computer Interaction}} \bibinfo{volume}{5}, \bibinfo{number}{CSCW2} (\bibinfo{year}{2021}), \bibinfo{pages}{1--41}.
\newblock


\bibitem[Shen et~al\mbox{.}(2021)]%
        {shen2021everyday}
\bibfield{author}{\bibinfo{person}{Hong Shen}, \bibinfo{person}{Alicia DeVos}, \bibinfo{person}{Motahhare Eslami}, {and} \bibinfo{person}{Kenneth Holstein}.} \bibinfo{year}{2021}\natexlab{}.
\newblock \showarticletitle{Everyday algorithm auditing: Understanding the power of everyday users in surfacing harmful algorithmic behaviors}.
\newblock \bibinfo{journal}{\emph{Proceedings of the ACM on Human-Computer Interaction}} \bibinfo{volume}{5}, \bibinfo{number}{CSCW2} (\bibinfo{year}{2021}), \bibinfo{pages}{1--29}.
\newblock


\bibitem[Siems et~al\mbox{.}(2022)]%
        {siems2022trade}
\bibfield{author}{\bibinfo{person}{Eli Siems}, \bibinfo{person}{Katherine~J Strandburg}, {and} \bibinfo{person}{Nicholas Vincent}.} \bibinfo{year}{2022}\natexlab{}.
\newblock \showarticletitle{Trade Secrecy and Innovation in Forensic Technology}.
\newblock \bibinfo{journal}{\emph{Hastings LJ}}  \bibinfo{volume}{73} (\bibinfo{year}{2022}), \bibinfo{pages}{773}.
\newblock


\bibitem[Spangenberg and Beeman(1995)]%
        {spangenberg1995indigent}
\bibfield{author}{\bibinfo{person}{Robert~L Spangenberg} {and} \bibinfo{person}{Marea~L Beeman}.} \bibinfo{year}{1995}\natexlab{}.
\newblock \showarticletitle{Indigent defense systems in the United States}.
\newblock \bibinfo{journal}{\emph{Law \& Contemp. Probs.}}  \bibinfo{volume}{58} (\bibinfo{year}{1995}), \bibinfo{pages}{31}.
\newblock


\bibitem[Suresh et~al\mbox{.}(2021)]%
        {suresh2021beyond}
\bibfield{author}{\bibinfo{person}{Harini Suresh}, \bibinfo{person}{Steven~R Gomez}, \bibinfo{person}{Kevin~K Nam}, {and} \bibinfo{person}{Arvind Satyanarayan}.} \bibinfo{year}{2021}\natexlab{}.
\newblock \showarticletitle{Beyond expertise and roles: A framework to characterize the stakeholders of interpretable machine learning and their needs}. In \bibinfo{booktitle}{\emph{Proceedings of the 2021 CHI Conference on Human Factors in Computing Systems}}. \bibinfo{pages}{1--16}.
\newblock


\bibitem[Suresh et~al\mbox{.}(2022)]%
        {suresh2022towards}
\bibfield{author}{\bibinfo{person}{Harini Suresh}, \bibinfo{person}{Rajiv Movva}, \bibinfo{person}{Amelia~Lee Dogan}, \bibinfo{person}{Rahul Bhargava}, \bibinfo{person}{Isadora Crux{\^e}n}, \bibinfo{person}{{\'A}ngeles~Martinez Cuba}, \bibinfo{person}{Guilia Taurino}, \bibinfo{person}{Wonyoung So}, {and} \bibinfo{person}{Catherine D'Ignazio}.} \bibinfo{year}{2022}\natexlab{}.
\newblock \showarticletitle{Towards intersectional feminist and participatory ML: A case study in supporting Feminicide Counterdata Collection}. In \bibinfo{booktitle}{\emph{Proceedings of the 2022 ACM Conference on Fairness, Accountability, and Transparency}}. \bibinfo{pages}{667--678}.
\newblock


\bibitem[Suresh et~al\mbox{.}(2023)]%
        {suresh2023kaleidoscope}
\bibfield{author}{\bibinfo{person}{Harini Suresh}, \bibinfo{person}{Divya Shanmugam}, \bibinfo{person}{Tiffany Chen}, \bibinfo{person}{Annie~G Bryan}, \bibinfo{person}{Alexander D'Amour}, \bibinfo{person}{John Guttag}, {and} \bibinfo{person}{Arvind Satyanarayan}.} \bibinfo{year}{2023}\natexlab{}.
\newblock \showarticletitle{Kaleidoscope: Semantically-grounded, context-specific ML model evaluation}. In \bibinfo{booktitle}{\emph{Proceedings of the 2023 CHI Conference on Human Factors in Computing Systems}}. \bibinfo{pages}{1--13}.
\newblock


\bibitem[Taori et~al\mbox{.}(2020)]%
        {taori2020measuring}
\bibfield{author}{\bibinfo{person}{Rohan Taori}, \bibinfo{person}{Achal Dave}, \bibinfo{person}{Vaishaal Shankar}, \bibinfo{person}{Nicholas Carlini}, \bibinfo{person}{Benjamin Recht}, {and} \bibinfo{person}{Ludwig Schmidt}.} \bibinfo{year}{2020}\natexlab{}.
\newblock \showarticletitle{Measuring robustness to natural distribution shifts in image classification}.
\newblock \bibinfo{journal}{\emph{Advances in Neural Information Processing Systems}}  \bibinfo{volume}{33} (\bibinfo{year}{2020}), \bibinfo{pages}{18583--18599}.
\newblock


\bibitem[Vaccaro et~al\mbox{.}(2019)]%
        {vaccaro2019contestability}
\bibfield{author}{\bibinfo{person}{Kristen Vaccaro}, \bibinfo{person}{Karrie Karahalios}, \bibinfo{person}{Deirdre~K Mulligan}, \bibinfo{person}{Daniel Kluttz}, {and} \bibinfo{person}{Tad Hirsch}.} \bibinfo{year}{2019}\natexlab{}.
\newblock \showarticletitle{Contestability in algorithmic systems}. In \bibinfo{booktitle}{\emph{Conference companion publication of the 2019 on computer supported cooperative work and social computing}}. \bibinfo{pages}{523--527}.
\newblock


\bibitem[Vaccaro et~al\mbox{.}(2020)]%
        {vaccaro2020end}
\bibfield{author}{\bibinfo{person}{Kristen Vaccaro}, \bibinfo{person}{Christian Sandvig}, {and} \bibinfo{person}{Karrie Karahalios}.} \bibinfo{year}{2020}\natexlab{}.
\newblock \showarticletitle{" At the End of the Day Facebook Does What ItWants" How Users Experience Contesting Algorithmic Content Moderation}.
\newblock \bibinfo{journal}{\emph{Proceedings of the ACM on human-computer interaction}} \bibinfo{volume}{4}, \bibinfo{number}{CSCW2} (\bibinfo{year}{2020}), \bibinfo{pages}{1--22}.
\newblock


\bibitem[Warren and Salehi(2022)]%
        {warren2022trial}
\bibfield{author}{\bibinfo{person}{Rachel~B Warren} {and} \bibinfo{person}{Niloufar Salehi}.} \bibinfo{year}{2022}\natexlab{}.
\newblock \showarticletitle{Trial by File Formats: Exploring Public Defenders' Challenges Working with Novel Surveillance Data}.
\newblock \bibinfo{journal}{\emph{Proceedings of the ACM on Human-Computer Interaction}} \bibinfo{volume}{6}, \bibinfo{number}{CSCW1} (\bibinfo{year}{2022}), \bibinfo{pages}{1--26}.
\newblock


\bibitem[Wexler(2017)]%
        {wexler2017computer}
\bibfield{author}{\bibinfo{person}{Rebecca Wexler}.} \bibinfo{year}{2017}\natexlab{}.
\newblock \showarticletitle{When a computer program keeps you in jail: How computers are harming criminal justice}.
\newblock \bibinfo{journal}{\emph{New York Times}}  \bibinfo{volume}{13} (\bibinfo{year}{2017}).
\newblock


\bibitem[Wexler(2018)]%
        {wexler2018life}
\bibfield{author}{\bibinfo{person}{Rebecca Wexler}.} \bibinfo{year}{2018}\natexlab{}.
\newblock \showarticletitle{Life, liberty, and trade secrets: Intellectual property in the criminal justice system}.
\newblock \bibinfo{journal}{\emph{Stan. L. Rev.}}  \bibinfo{volume}{70} (\bibinfo{year}{2018}), \bibinfo{pages}{1343}.
\newblock


\bibitem[Wilson-Kovacs et~al\mbox{.}(2023)]%
        {wilson2023digital}
\bibfield{author}{\bibinfo{person}{Dana Wilson-Kovacs}, \bibinfo{person}{Rebecca Helm}, \bibinfo{person}{Beth Growns}, {and} \bibinfo{person}{Lauren Redfern}.} \bibinfo{year}{2023}\natexlab{}.
\newblock \showarticletitle{Digital evidence in defence practice: Prevalence, challenges and expertise}.
\newblock \bibinfo{journal}{\emph{The International Journal of Evidence \& Proof}} (\bibinfo{year}{2023}), \bibinfo{pages}{13657127231171620}.
\newblock


\bibitem[Wright et~al\mbox{.}({[n.\,d.]})]%
        {wright2024null}
\bibfield{author}{\bibinfo{person}{Lucas Wright}, \bibinfo{person}{Roxana~Mika Muenster}, \bibinfo{person}{Briana Vecchione}, \bibinfo{person}{Tianyao Qu}, \bibinfo{person}{Senhuang~(Pika) Cai}, \bibinfo{person}{Alan Smith}, \bibinfo{person}{Jacob Metcalf}, {and} \bibinfo{person}{J.~Nathan Matias}.} \bibinfo{year}{[n.\,d.]}\natexlab{}.
\newblock \showarticletitle{Null Compliance: NYC Local Law 144 and the Challenges of Algorithm Accountability}.
\newblock  (\bibinfo{year}{[n.\,d.]}).
\newblock
\urldef\tempurl%
\url{https://doi.org/10.17605/OSF.IO/UPFDK}
\showURL{%
\tempurl}


\bibitem[Wu et~al\mbox{.}(2021)]%
        {wu2021medical}
\bibfield{author}{\bibinfo{person}{Eric Wu}, \bibinfo{person}{Kevin Wu}, \bibinfo{person}{Roxana Daneshjou}, \bibinfo{person}{David Ouyang}, \bibinfo{person}{Daniel~E Ho}, {and} \bibinfo{person}{James Zou}.} \bibinfo{year}{2021}\natexlab{}.
\newblock \showarticletitle{How medical AI devices are evaluated: limitations and recommendations from an analysis of FDA approvals}.
\newblock \bibinfo{journal}{\emph{Nature Medicine}} \bibinfo{volume}{27}, \bibinfo{number}{4} (\bibinfo{year}{2021}), \bibinfo{pages}{582--584}.
\newblock


\bibitem[Yang et~al\mbox{.}(2023a)]%
        {yang2023subjective}
\bibfield{author}{\bibinfo{person}{Fumeng Yang}, \bibinfo{person}{Maryam Hedayati}, {and} \bibinfo{person}{Matthew Kay}.} \bibinfo{year}{2023}\natexlab{a}.
\newblock \showarticletitle{Subjective Probability Correction for Uncertainty Representations}. In \bibinfo{booktitle}{\emph{Proceedings of the 2023 CHI Conference on Human Factors in Computing Systems}}. \bibinfo{pages}{1--17}.
\newblock


\bibitem[Yang et~al\mbox{.}(2023b)]%
        {yang2023designing}
\bibfield{author}{\bibinfo{person}{Qian Yang}, \bibinfo{person}{Richmond~Y Wong}, \bibinfo{person}{Thomas Gilbert}, \bibinfo{person}{Margaret~D Hagan}, \bibinfo{person}{Steven Jackson}, \bibinfo{person}{Sabine Junginger}, {and} \bibinfo{person}{John Zimmerman}.} \bibinfo{year}{2023}\natexlab{b}.
\newblock \showarticletitle{Designing Technology and Policy Simultaneously: Towards A Research Agenda and New Practice}. In \bibinfo{booktitle}{\emph{Extended Abstracts of the 2023 CHI Conference on Human Factors in Computing Systems}}. \bibinfo{pages}{1--6}.
\newblock


\bibitem[Yin et~al\mbox{.}(2019)]%
        {yin2019understanding}
\bibfield{author}{\bibinfo{person}{Ming Yin}, \bibinfo{person}{Jennifer Wortman~Vaughan}, {and} \bibinfo{person}{Hanna Wallach}.} \bibinfo{year}{2019}\natexlab{}.
\newblock \showarticletitle{Understanding the effect of accuracy on trust in machine learning models}. In \bibinfo{booktitle}{\emph{Proceedings of the 2019 chi conference on human factors in computing systems}}. \bibinfo{pages}{1--12}.
\newblock


\bibitem[Yurrita et~al\mbox{.}(2023)]%
        {yurrita2023disentangling}
\bibfield{author}{\bibinfo{person}{Mireia Yurrita}, \bibinfo{person}{Tim Draws}, \bibinfo{person}{Agathe Balayn}, \bibinfo{person}{Dave Murray-Rust}, \bibinfo{person}{Nava Tintarev}, {and} \bibinfo{person}{Alessandro Bozzon}.} \bibinfo{year}{2023}\natexlab{}.
\newblock \showarticletitle{Disentangling Fairness Perceptions in Algorithmic Decision-Making: the Effects of Explanations, Human Oversight, and Contestability}. In \bibinfo{booktitle}{\emph{Proceedings of the 2023 CHI Conference on Human Factors in Computing Systems}}. \bibinfo{pages}{1--21}.
\newblock


\bibitem[Zhang et~al\mbox{.}(2020)]%
        {zhang2020effect}
\bibfield{author}{\bibinfo{person}{Yunfeng Zhang}, \bibinfo{person}{Q~Vera Liao}, {and} \bibinfo{person}{Rachel~KE Bellamy}.} \bibinfo{year}{2020}\natexlab{}.
\newblock \showarticletitle{Effect of confidence and explanation on accuracy and trust calibration in AI-assisted decision making}. In \bibinfo{booktitle}{\emph{Proceedings of the 2020 conference on fairness, accountability, and transparency}}. \bibinfo{pages}{295--305}.
\newblock


\end{thebibliography}

\onecolumn
\appendix
\section{Interview storyboards}
\label{appendix}
\begin{figure*}[ht!]
    \centering
    \includegraphics[width=\textwidth]{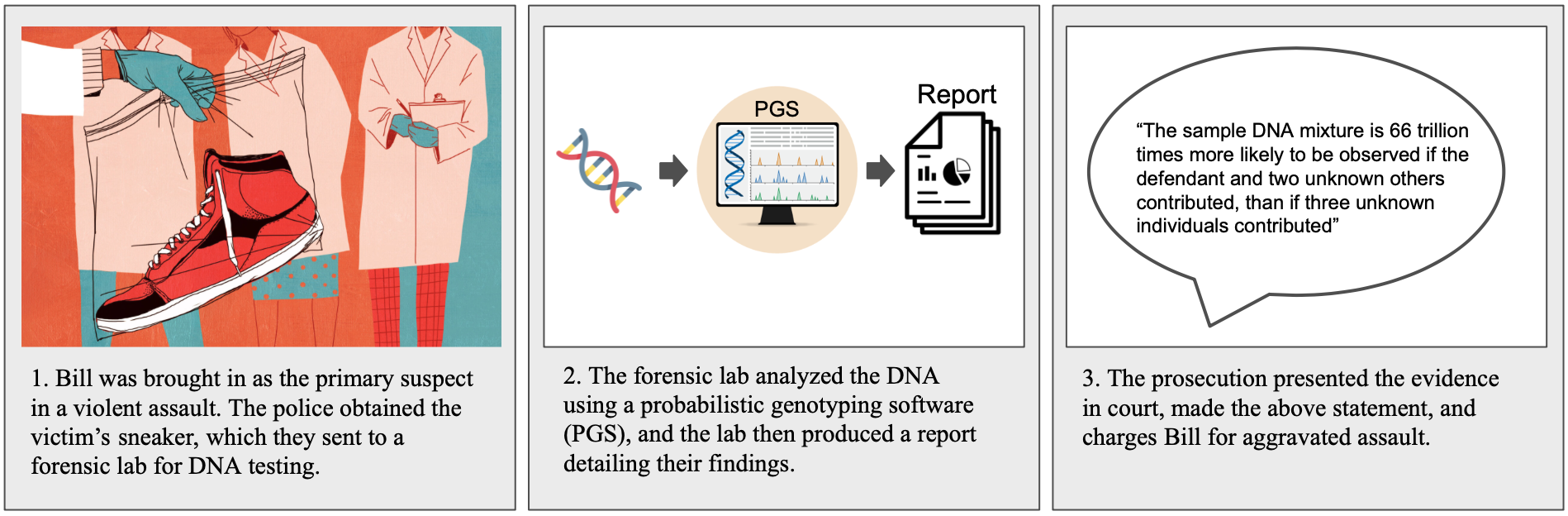}
    \caption{We used this storyboard to establish a hypothetical scenario in which the prosecution uses evidence output by probabilistic genotyping software (PGS). We presented this scenario before presenting potential evaluation approaches depicted in Figures~\ref{fig:adv-scrut-storyboard},~\ref{fig:checklist-storyboard}, and~\ref{fig:model-cards-storyboard}. This scenario is modeled off of a real case publicly documented by~\citet{kirchner2017thousands}, and uses an image from the article created by Michael Hirshon for ProPublica.}.
    \label{fig:pgs-scenario-intro}
    \Description{Scenes depicting police obtaining DNA evidence from a crime scene, a forensic lab analyzing the DNA evidence using probabilistic genotyping software to produce a report, and the prosecution presenting the evidence in court.}
\end{figure*}

\begin{figure*}[ht!]
    \centering
    \includegraphics[width=\textwidth]{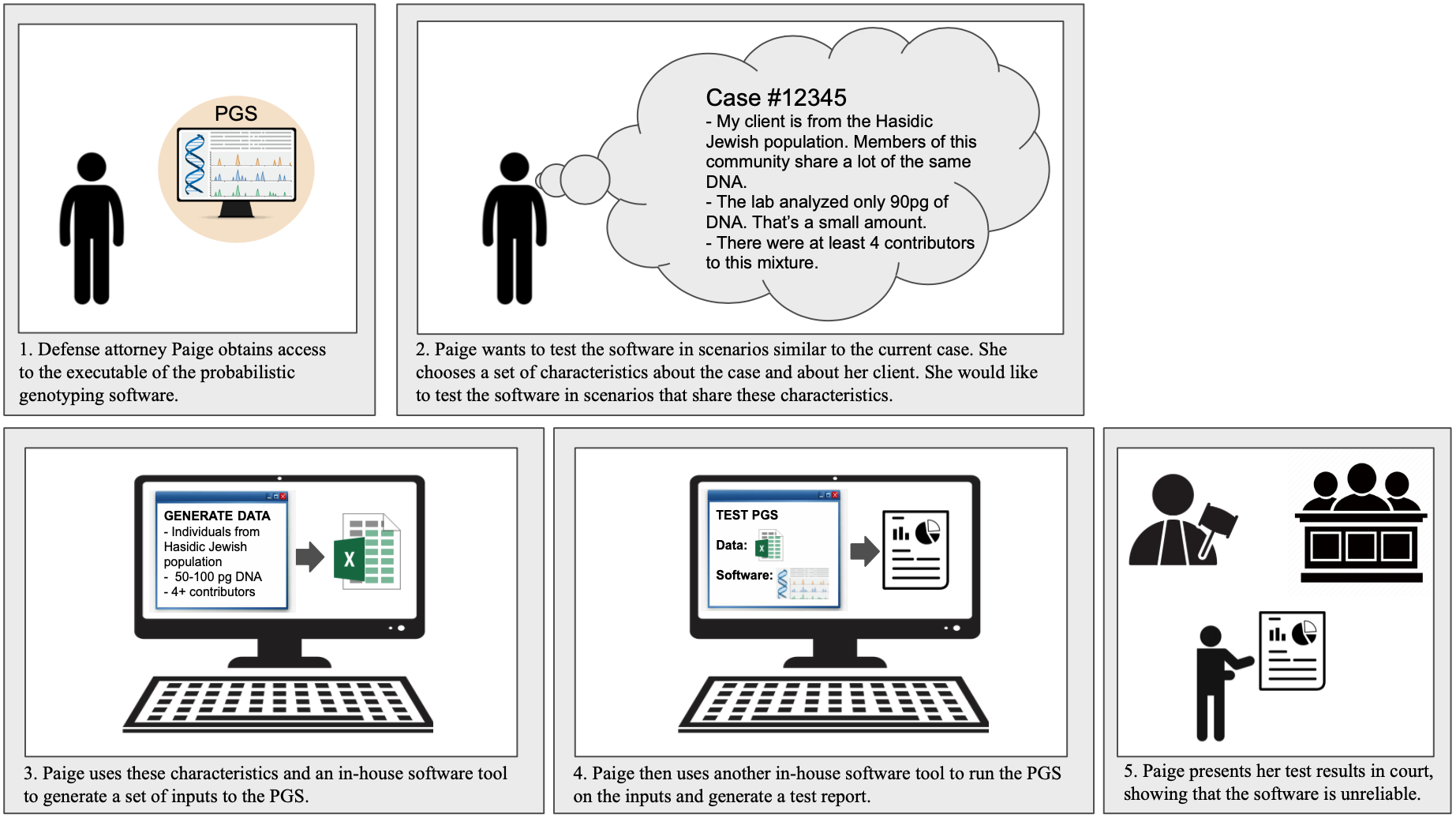}
    \caption{This storyboard depicts a potential evaluation approach based on~\citet{abebe2022adversarial}.}
    \label{fig:adv-scrut-storyboard}
    \Description{Scenes depicting a defense attorney obtaining executable access to the PGS, deciding on how to test the software for the specific case at hand, generating a set of inputs to test the PGS, running the PGS system on the inputs and generating a test report, and presenting her results in court to show that the software is unreliable.}
\end{figure*}

\begin{figure*}[ht!]
    \centering
    \includegraphics[width=\textwidth]{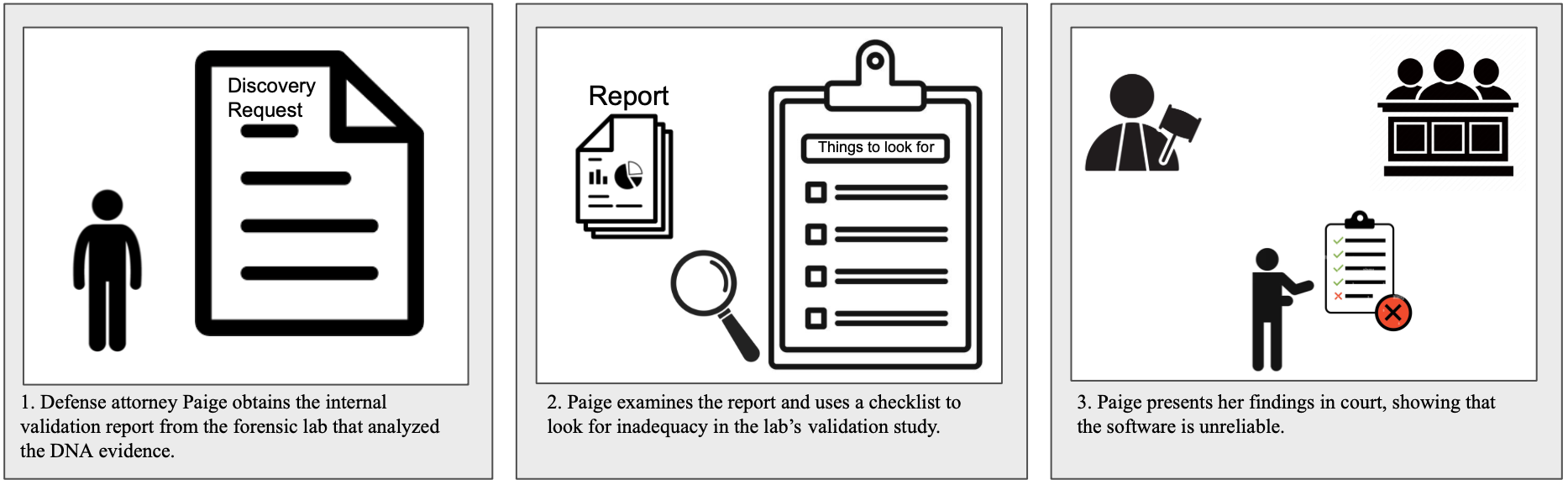}
    \caption{This storyboard depicts a potential evaluation approach based on the Justice in Forensic Algorithms Act of 2021, a bill introduced to Congress by Rep. Mark Takano [D-CA-41]~\cite{takano2021}.}
    \label{fig:checklist-storyboard}
    \Description{Scenes depicting a defense attorney obtaining an internal validation report from the forensic lab that analyzed the evidence, examining the report using a checklist to identify inadequacies in the lab's validation study, and presenting findings from the analysis in court to show that the software is unreliable.}
\end{figure*}

\begin{figure*}[ht!]
    \centering
    \includegraphics[width=\textwidth]{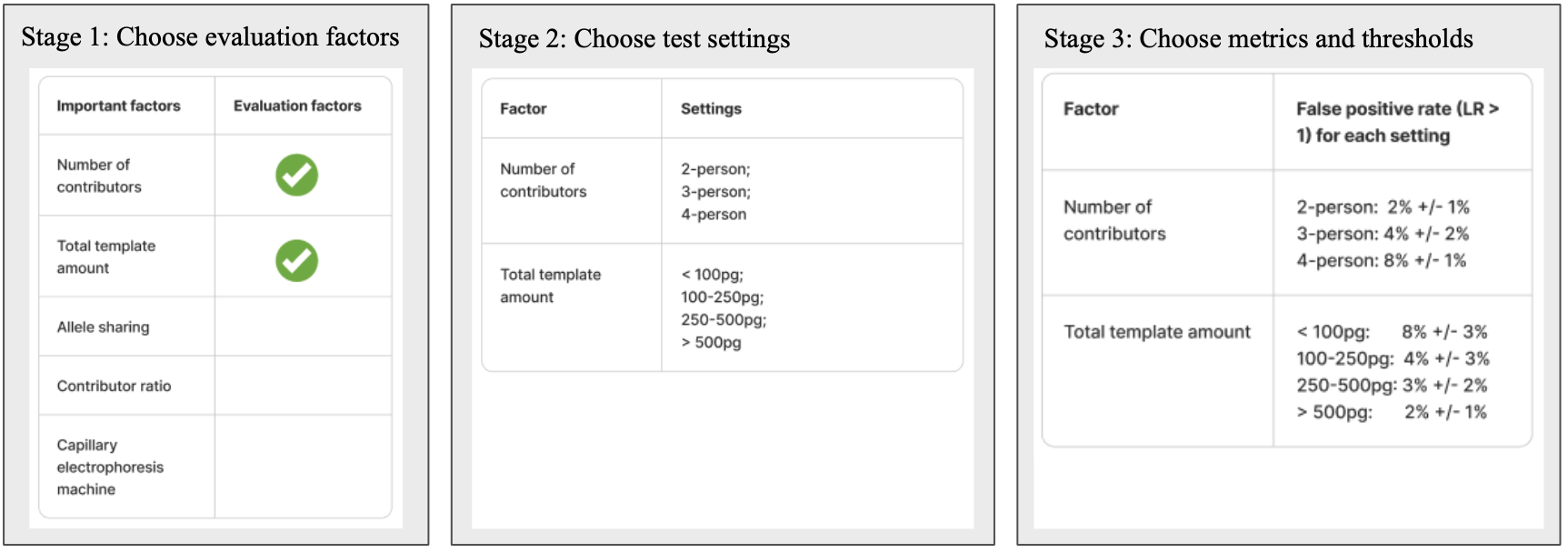}
    \caption{This storyboard depicts a potential evaluation approach based on~\citet{mitchell2019model}.}
    \label{fig:model-cards-storyboard}
    \Description{Scenes depicting steps an attorney might take in conducting a traditional performance evaluation of PGS. The first step involves choosing a set of factors to evaluate. The next step involves choosing test settings for each factor. The last step involves choosing metrics and thresholds for the test.}
\end{figure*}

We include the storyboards we developed and used for the study in Figures~\ref{fig:pgs-scenario-intro},~\ref{fig:adv-scrut-storyboard},~\ref{fig:checklist-storyboard}, and~\ref{fig:model-cards-storyboard}.

\end{document}